\documentclass[twocolumn,prd,superscriptaddress,preprintnumbers,amsmath,amssymb,showpacs,nofootinbib,noshowpacs]{revtex4}
\usepackage{mathbbol}

\usepackage{color}
\usepackage{setspace}
\usepackage{latexsym}
\usepackage{verbatim}

\usepackage[pdftex]{graphicx}

\usepackage{subfigure}
\usepackage{rotating}
\usepackage{bm}
\usepackage{tikz}

\definecolor{myred}{rgb}{1.00,0,0}

\definecolor{myblue}{rgb}{0,0,1.00}

\usepackage{ulem}
\renewcommand{\emph}[1]{{\tt #1}}

\newlength{\twopicwidth} 
\newlength{\threepicwidth} 

\newcounter{ecutleft} 
\newcounter{ecutright} 
\newcounter{ecutbottom} 
\newcounter{ecuttop} 

\newcounter{dcutleft} 
\newcounter{dcutright} 
\newcounter{dcutbottom} 
\newcounter{dcuttop} 

\newcounter{scutleft} 
\newcounter{scutright} 
\newcounter{scutbottom} 
\newcounter{scuttop} 

\def\fig{Fig.~}
\def\Fig{Fig.~}

\def\eqn{Eq.~}

\newcommand{\ihat}{\hat{\imath}}

\begin{document}

\title{Anderson localization through Polyakov loops:\\ lattice evidence and Random matrix model}
\author{Falk Bruckmann}\affiliation{Institut f\"ur Theoretische Physik, D-93040 Regensburg, Germany}
\author{Tam\'as G.\ Kov\'acs}\affiliation{Department of Physics, University of P\'ecs, H-7624 P\'ecs, Ifj\'us\'ag \'utja 6, Hungary}
\author{Sebastian Schierenberg}\affiliation{Institut f\"ur Theoretische Physik, D-93040 Regensburg, Germany}


\begin{abstract}
 We investigate low-lying fermion modes in $SU(2)$ gauge theory at
 temperatures above the phase transition. Both staggered and overlap spectra reveal transitions
 from chaotic (random matrix) to integrable (Poissonian) behavior
 accompanied by an increasing localization
 of the eigenmodes. We show that the latter are trapped by local Polyakov loop
 fluctuations. Islands of such ``wrong'' Polyakov loops can therefore be
 viewed as defects leading to Anderson localization in gauge theories.
 We find strong similarities in the spatial profile of these localized
 staggered and overlap eigenmodes.
 We discuss possible
 interpretations of this finding and present a sparse random matrix model that
 reproduces these features.

\end{abstract}

\maketitle

\section{Introduction}

The vacuum of Quantum Chromodynamics (QCD) is a prominent nonperturbative
system, whose strongly interacting nature persists even above the transition
to the quark-gluon plasma.  To get insight into its mechanism, spectral
properties of the QCD Dirac operator are very useful.  A nonzero density of
eigenvalues at zero gives rise to chiral symmetry breaking via the
Banks-Casher formula \cite{Banks:1980}.  Moreover, exact zero modes are
related to the topological charge via index theorems.

In recent years, localization properties of the Dirac eigenmodes have
attracted attention as they can be used to draw analogies to condensed matter
phenomena: concepts like the mobility edge and Anderson localization can be
studied in QCD lattice simulations. In this spirit the chiral transition at
finite temperature\footnote{At zero temperature the situation is not clear due
  to the continuum limit, see \cite{Forcrand:2007a} and references therein.}
has been conjectured to be an Anderson (metal-insulator) transition.

This goes hand in hand with different random matrix theory (RMT) descriptions
of the Dirac spectra. In the low temperature phase, the existence of the
chiral condensate connects QCD to chiral perturbation theory and random matrix
theory (becoming exact in the epsilon-regime), which explains the statistics
of the low lying part of the Dirac spectrum \cite{Verbaarschot:2000}.

The spectral gap in the high temperature phase, on the other hand, seems to
call for the ``soft edge'' description of RMT, which, however, could not be
supported by lattice data \cite{Farchioni:2000,Damgaard:2000,Narayanan:2006sd,Narayanan:2006ek}. Instead, a
transition to independent eigenmodes obeying Poisson statistics has been
suggested \cite{Garcia-Garcia:2007}. A refined analysis by one of us has
shown, that the bulk of the spectrum is still delocalized and subject to RMT,
while the lowest lying eigenmodes display a transition to localization and
Poissonian behavior of the eigenvalues \cite{Kovacs:2010,Kovacs:2010a}. This
effect has been confirmed to be universal in the sense that it does not depend
on the resolution of the lattice. Observables, like e.g.\ the number of
localized modes, rather scale with the physical volume. On the other hand,
Ref.\ \cite{Gavai:2008} found that localization is essentially a finite volume
artifact. However, they used a different, less restrictive definition of
localization and thus their results are not in conflict with the rest of the
above cited literature.

In this work we give another crucial ingredient of Anderson localization in
QCD, namely we identify the ``defects'' causing it. We show that at high
temperature local Polyakov loop fluctuations trap low-lying modes. The
average Polyakov loop as an order parameter of the deconfinement (or center)
phase transition approaches 1 with increasing temperature.
Locally, however, the Polyakov loop takes on other values, in
particular close to other center elements, $-1$ in the case of gauge group $SU(2)$. The phase transition can actually
be viewed as a percolation of the physical sector embracing such islands
where the Polyakov loop is close to other center elements
\cite{Yaffe:1982,Gattringer:2010}.

The locations of these Polyakov loop fluctuations (which are similar to Weiss
domains in ferromagnetism) are correlated with maxima in the profile of
low-lying Dirac modes. We will demonstrate this for eigenmodes of both the
overlap and staggered operator. Similarities between the spectrum of the
overlap and staggered Dirac operator have already been found in the Schwinger
model \cite{Durr:2003xs} and also in QCD \cite{Durr:2004as}.  Our present
study is, however, the first one when similarity of staggered and overlap
Dirac eigen{\it modes} is seen in lattice simulations and we consider this an
important side result of our study.

The observed localization can be understood via Polyakov loops compensating the
twist caused by the antiperiodic boundary conditions in the temporal
direction. The corresponding Matsubara frequency is effectively lowered
(locally) which results in lower eigenvalues. Actually, our finding has been
inspired by a similar localization effect in the spectrum of the
gauge-covariant Laplace operator \cite{Bruckmann:2005c}. The Laplacian is the
square of the Dirac operator in the free case, so the twist picture
applies. Otherwise this operator does not share important chiral features like
topological zero modes and condensates.

Likewise, our finding is consistent with the existence of a chiral condensate
in the Polyakov loop sector close to other center elements
\cite{Chandrasekharan:1995gt,Stephanov:1996he,Meisinger:1995ih,Chandrasekharan:1995nf,Gattringer:2002dv,Gattringer:2002tg,Bornyakov:2008bg,Kovacs:2008sc}
(also needed for center symmetry breaking \cite{Bilgici:2008}), which implies
low Dirac eigenvalues in islands of such Polyakov loops.

The connection of these islands to topological excitations like magnetic
monopoles is attractive, but in its naive form contradicts the observed
topological susceptibility quantitatively, see below.

For the construction of random matrix models valid at high temperature we
investigate the distribution of local Polyakov loops and find them to be
uncorrelated to a good approximation. Hence, Polyakov loops in fact provide
the Poissonian ingredient for the Dirac spectrum.  This is built into a novel
Anderson-like random matrix model through supplementing it by random matrix
entries that represent nearest neighbor hoppings in three-dimensional space. We motivate this
model and show, that with a few parameters it reproduces the main features of
the Dirac modes: chirality, spectral gap, RMT-Poisson transition and
localization (to the analogue of local Polyakov loops).

Our findings are based on quenched lattice simulations with the SU(2) gauge group, we strongly believe that the described phenomena are present in more realistic gauge theories, too.\\

The paper is organized as follows. In the next section we describe the Dirac
spectra at high temperature including the RMT-Poisson (chaotic to integrable)
transition and the similarity of the staggered and overlap
modes. Sect.~\ref{sect polyakov} is devoted to the connection between local
Polyakov loops and low-lying modes. In Sect.~\ref{sect interpretation}
thereafter we investigate two possible interpretations of this finding,
effective Matsubara frequencies and topological objects. In  Section
\ref{sect rmt} we introduce and explore our 
random matrix model and finally Section \ref{sect conc} contains our conclusions.

\begin{figure}[b]
\includegraphics[width=1.05\linewidth]{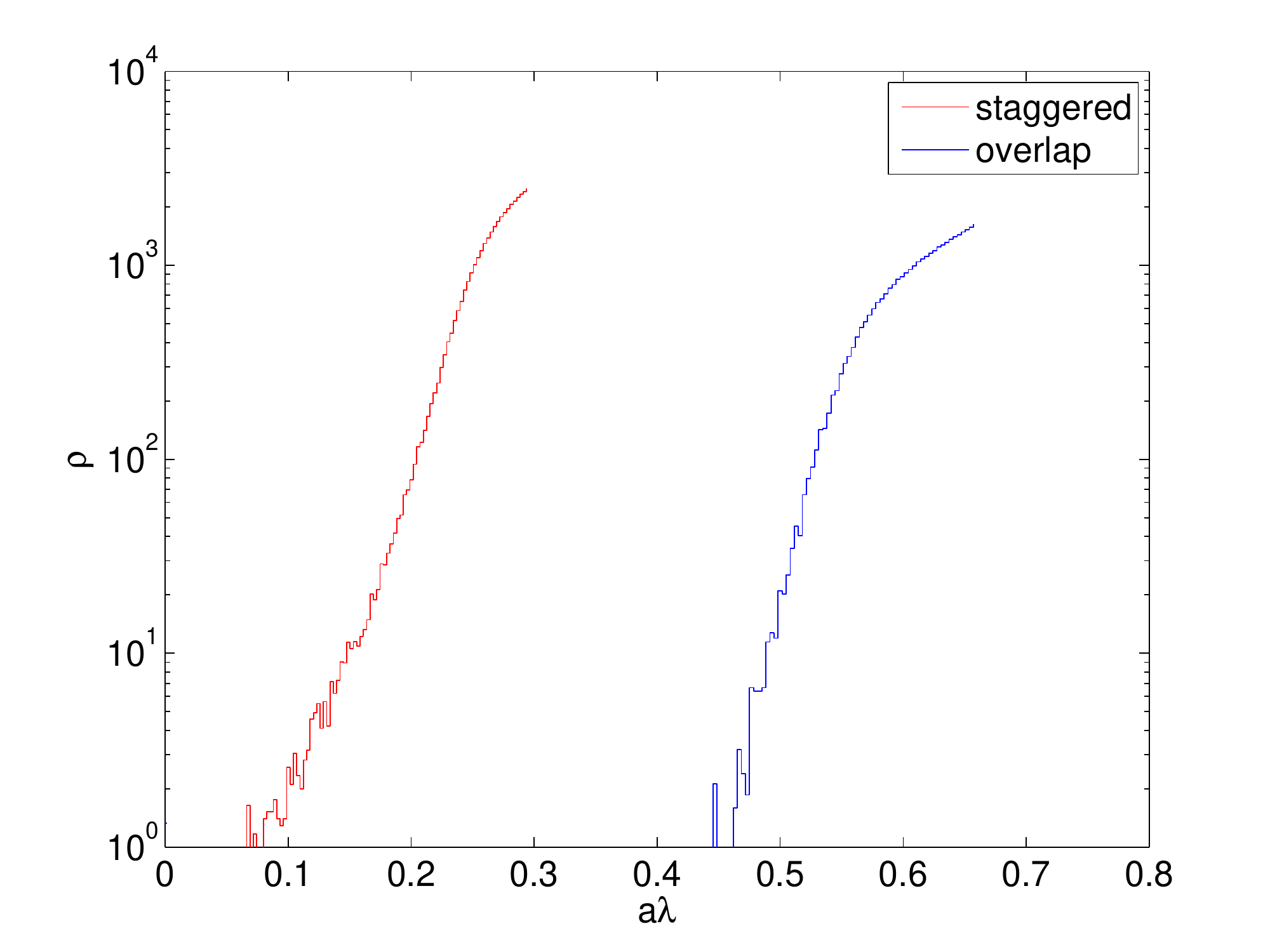}
\caption{Logarithm of the spectral density along the imaginary axis for
  the staggered (red, at smaller $\lambda$) and overlap (blue) Dirac operator
  from the lowest $256$ eigenvalues.}
\label{fig spec density}
\end{figure}

\section{Dirac spectra at high temperature}
\label{sect dirac}

We analyze quenched SU(2) lattice configurations generated with Wilson action
on a $24^3\cdot 4$ lattice at $\beta=2.6$ which amounts to a temperature of
$2.6 T_c$. The average Polyakov loop of $\,0.37\,$ signals deconfinement (by
Polyakov loop we refer to the trace of the products of all temporal links
$L(\vec{x})=1/2\cdot {\rm Tr}\prod_{x_0=1}^{N_t}U_0(x_0,\vec{x})$ and we selected the physical sector of positive Polyakov loops by hand).

We measured the $256$ lowest eigenvalues with positive imaginary parts
of the overlap \cite{Neuberger:1997fp,Neuberger:1998wv} (with pa\-ra\-me\-ter $s=0.4$ cf.\ \cite{Kovacs:2010}) and staggered
Dirac operator on 1136 (overlap)/
3149 (staggered) configurations. For a set of 1102 configurations, we also
measured the 12 lowest eigenmodes of both staggered and overlap
operator. In all cases the quark mass was set to zero.

The eigenvalues are ordered according to their imaginary parts and the corresponding
eigenvalue densities\footnote{Due to chirality, the nonzero eigenvalues come
in pairs of opposite imaginary part and we restrict ourselves to the half
with positive imaginary part. In other words, all plots can be extended
symmetrically around $\lambda=0$.} are plotted in \fig\ref{fig spec
density}.  They display a gap-like behavior with the eigenvalue density
starting to differ from zero considerably at $a\lambda\simeq 0.15$ and $a\lambda\simeq 0.5$,
respectively. In addition, the overlap operator possesses exact zero modes, which we use to
determine the topological charge of the configuration.

To describe the RMT vs.\ Poissonian behavior of the Dirac spectra we will scan
windows in the range of available eigenvalues and measure the level spacing
distributions $P(s)$ on unfolded eigenvalues \cite{Verbaarschot:2000}, a
typical quantity describing the eigenvalue statistics. The Gaussian RMT ensembles
provide predictions for it, which also apply to systems with chiral symmetry,
depending only on the universality class.  For gauge group
$SU(2)$ the latter are the Gaussian Orthogonal Ensemble (GOE) for the overlap operator
(like for the continuum Dirac operator) and the Gaussian Symplectic Ensemble (GSE) for
the staggered operator. The main difference of those $P(s)$ formulae lies in
the different repulsion strength of nearby eigenvalues, which results in
a linear and quartic behavior of $P(s)$ near $s=0$,
respectively. Independent eigenvalues, on the
other hand, lack such a repulsion and the unfolded level spacing is a Poissonian distribution, i.e.\ $P(s)=\exp(-s)$.

\begin{figure*}[!t]
\includegraphics[height=1.1\threepicwidth]{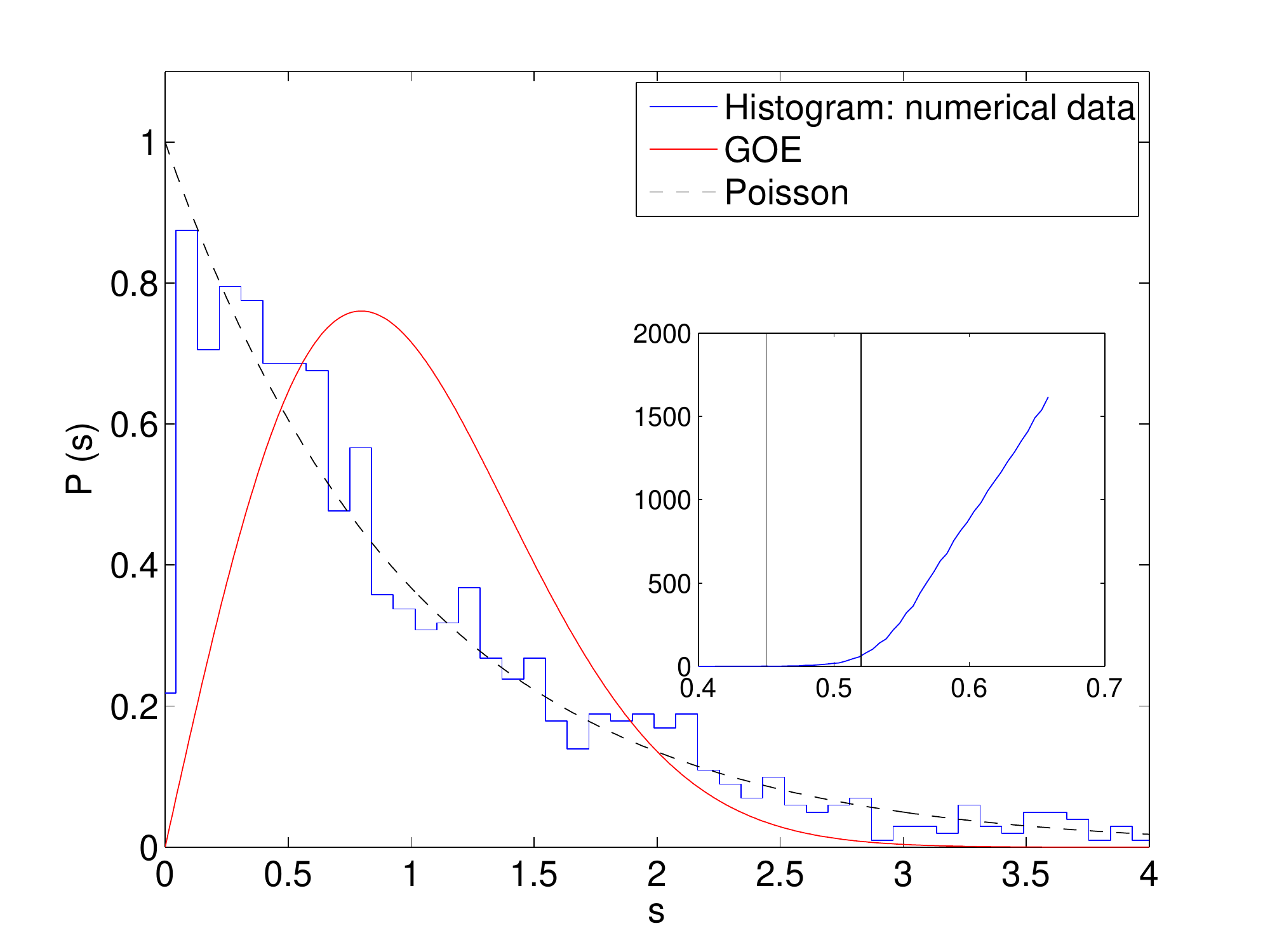}\hspace*{-0.9cm}
\includegraphics[height=1.1\threepicwidth]{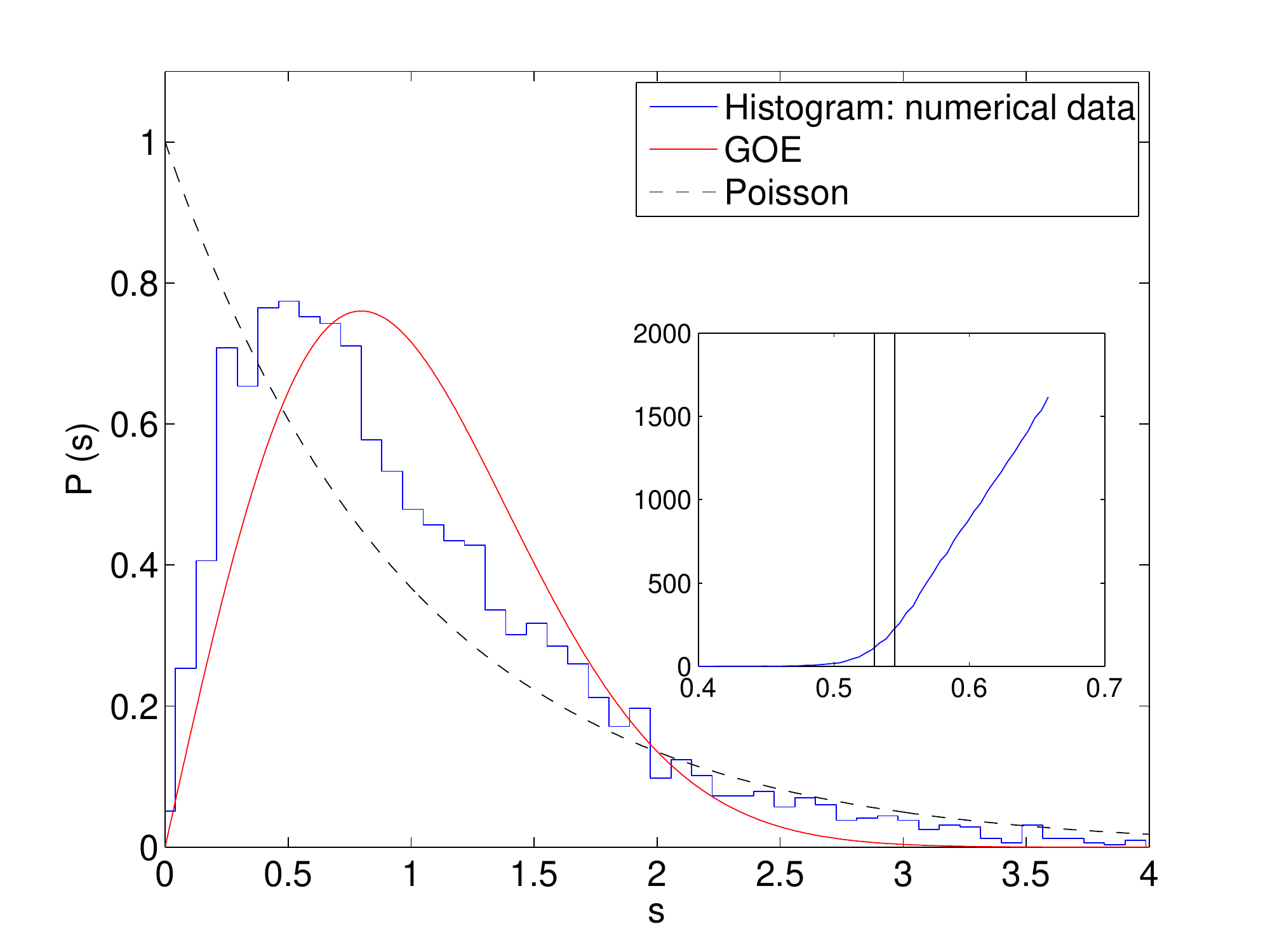}\hspace*{-0.9cm}
\includegraphics[height=1.1\threepicwidth]{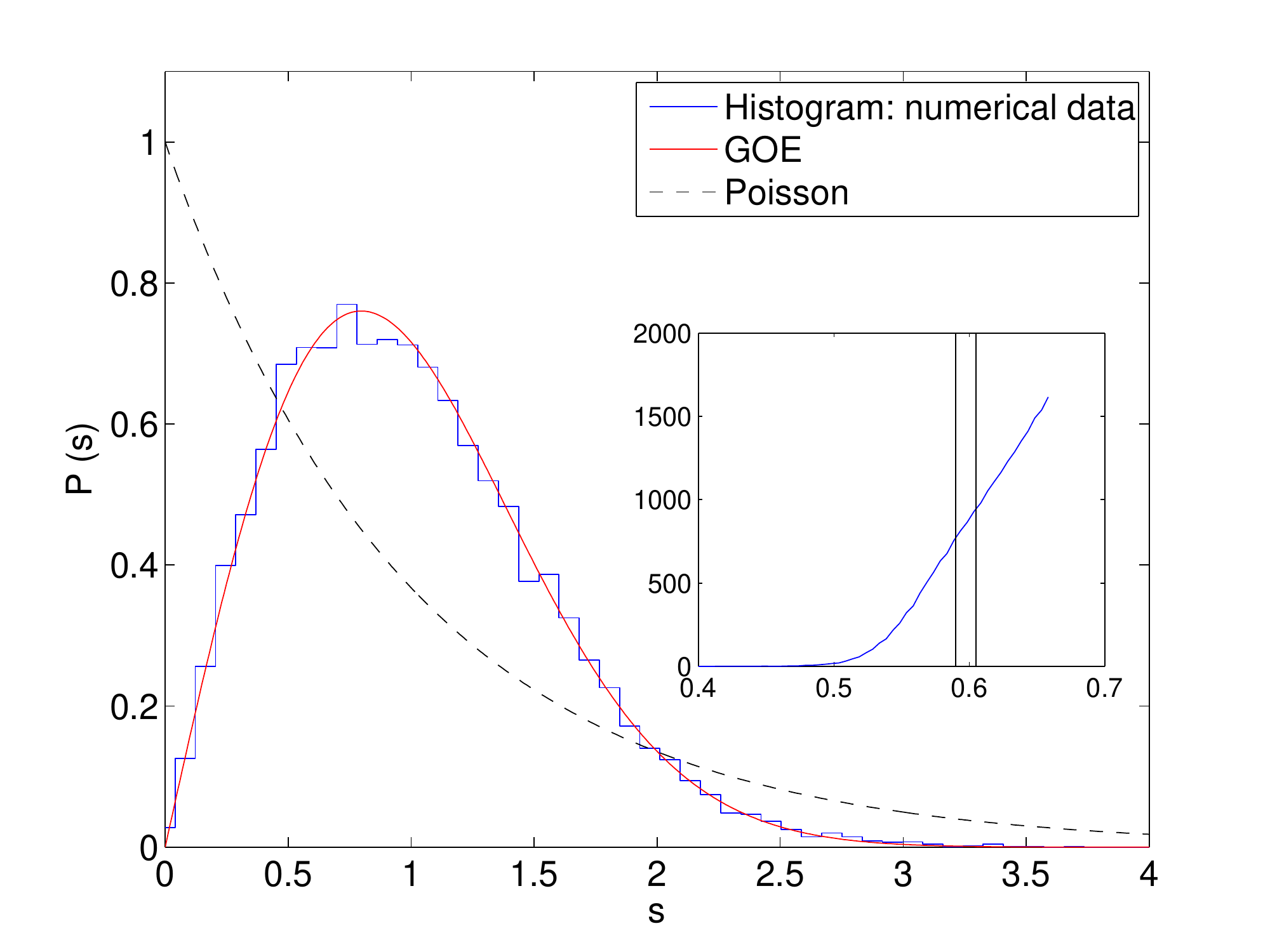}
\caption{Spacing distributions of the overlap spectrum in spectral windows
  indicated by the insets showing the spectral density.  The pure RMT
  (GOE) prediction and the Poissonian distribution are plotted for comparison.}
\label{fig transition overlap}
\end{figure*}

\Fig\ref{fig transition overlap} shows the level spacing distribution of overlap eigenvalues (see \cite{Kovacs:2010a} for staggered spectra). It 
clearly reveals that the level spacing agrees with the associated RMT predictions in the bulk and moves
towards Poissonian when the spectral window is shifted towards lower eigenvalues. To observe such a
level {\it spacing} at least a few
  independently (Poissonian) distributed eigenvalues are needed on each
  configuration. Since independent modes occur only at the very low end of the
  spectrum where the spectral density is low, large enough volumes are required
  for that.  In \cite{Kovacs:2010} also the independence of these data of the
lattice spacing has been demonstrated for the staggered case. 

 The properties of the independent and localized modes at the lower ends of
 the spectra are the main subject of the rest of the paper. We therefore check
 first, whether the modes of the overlap and the staggered operator see
 similar physical effects, meaning that their profiles are correlated and
 localized to similar locations. 

\begin{figure}[b]
\includegraphics[width=0.9\linewidth]{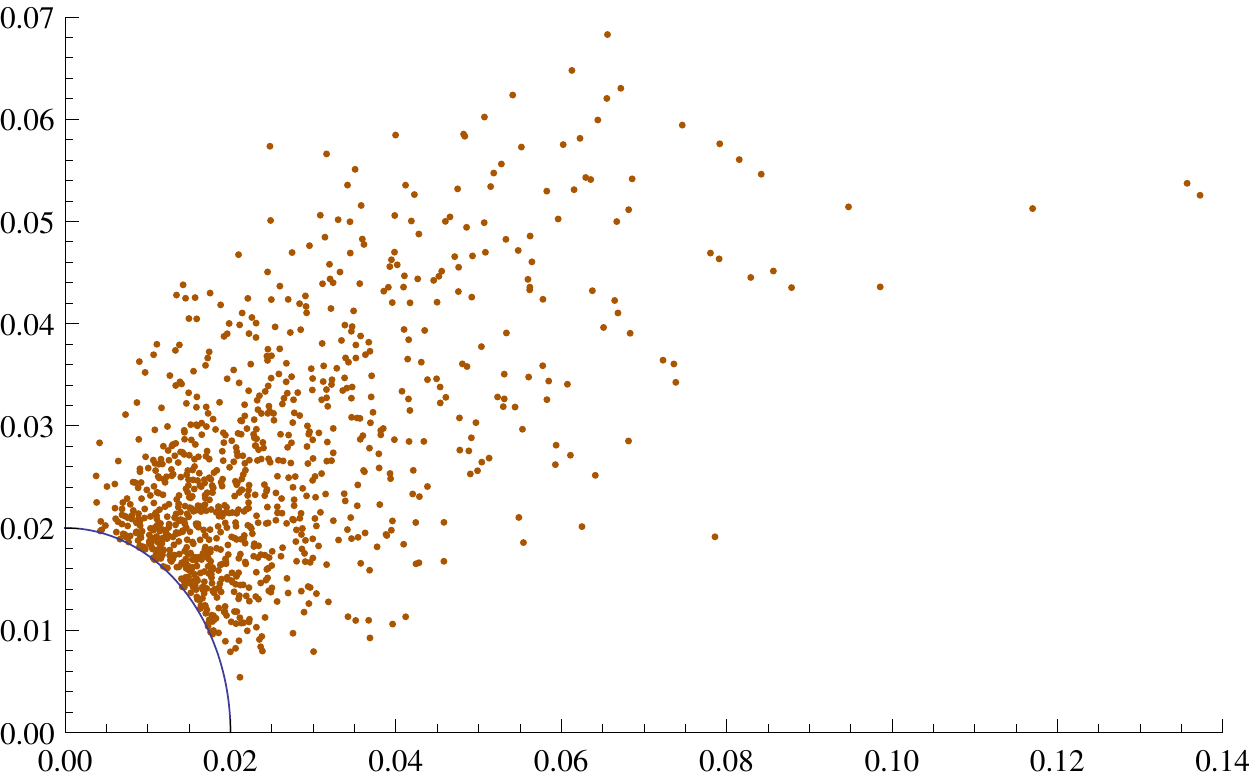}
\caption{Scatter plot of staggered (horizontal) vs.\ overlap (vertical)
  amplitudes for the lowest modes, $|\psi^{{\rm st}}_1(x)|$ vs. $|\psi^{{\rm ov}}_1(x)|$, 
  on a $Q=0$ configuration over the whole lattice. Data points with small amplitudes -- as indicated by the circle -- have been excluded to avoid overcrowding the plot.}
\label{fig scatter overlap staggered}
\end{figure}

First of all we remark that for every overlap zero mode we find a staggered
eigenvalue with unusually small value. In the topological sector $Q=0$, the
average smallest eigenvalue is 0.175(1), whereas it is 0.109(3) for $|Q|=1$
(and 0.098(8)/0.141(6) for $|Q| = 2$, where we have two small
eigenvalues). This can also be seen in \fig\ref{fig polloop modes one}
bottom.

\begin{figure}[b]
\includegraphics[width=0.9\linewidth]{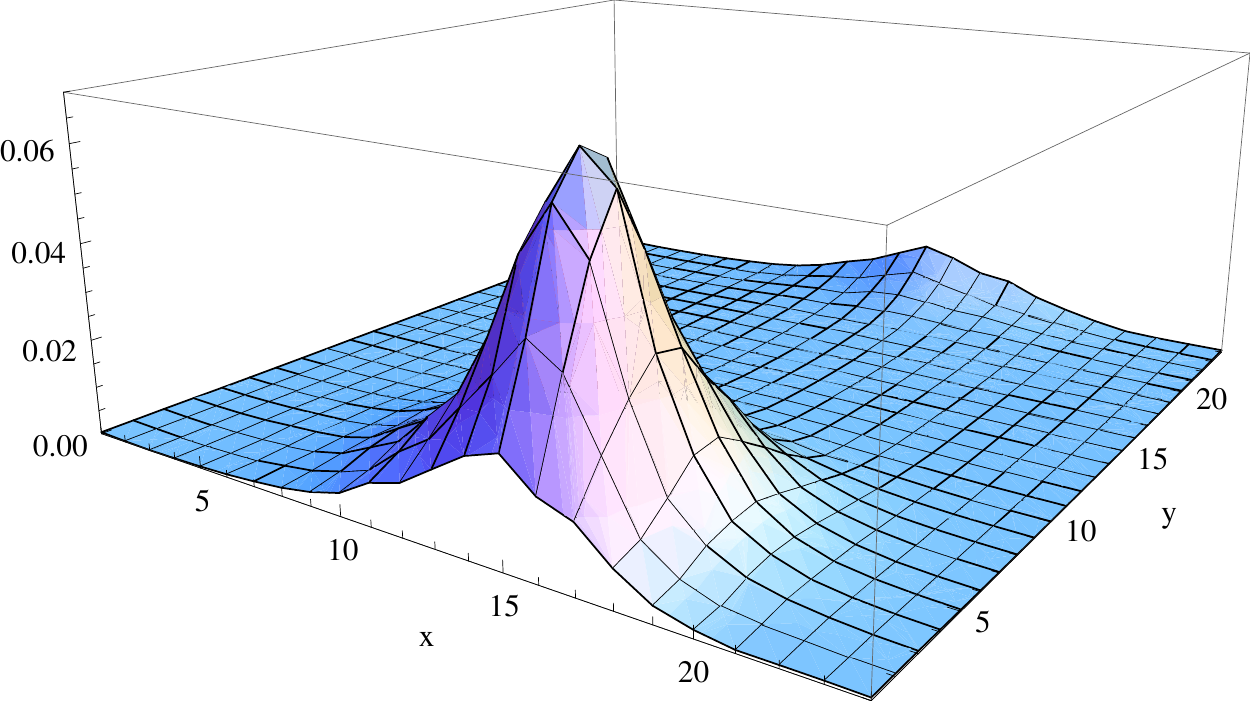}\\
\includegraphics[width=0.9\linewidth]{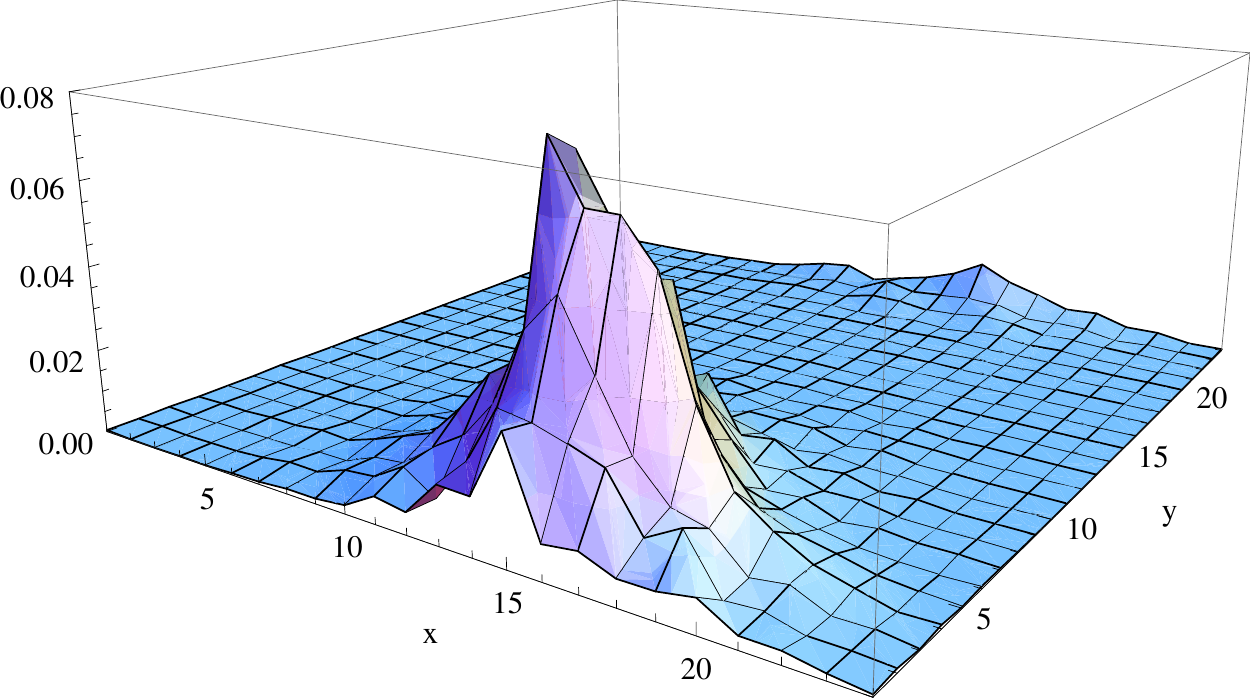}
\caption{Profile of the overlap (top) and staggered (bottom) lowest mode 
of the configuration of Fig.~\protect\ref{fig scatter overlap staggered},
in a lattice plane where the overlap mode takes on its maximum.
(The absolute maximum of the staggered mode is separated from the overlap one by $\sqrt{2}$ lattice spacings in the remaining directions.)}
\label{fig profile modes}
\end{figure}

Next, the scatter plot of \fig\ref{fig scatter overlap staggered} gives a strong indication
for the correlation of the local mode amplitudes in a typical example configuration, which
is further visualized by the two-dimensional profiles in \fig\ref{fig profile modes}.

In order to quantify the similarity and localization of two modes we propose the following ``interlocalization''
\begin{equation}
I := N\sum_{x} \left|\psi_m^{{\rm ov}}(x)\right|^2\left|\psi_n^{{\rm st}}(x)\right|^2\,,
\label{eqn interlocalisation}
\end{equation}
where $N$ is the number of lattice sites, and $|\psi_m^{{\rm ov,st}}(x)|$ is
the absolute value ($L^2$-norm) of the $m$th overlap/staggered eigenmode summed up
over gauge -- and in the case of overlap also spinor -- indices at lattice site $x$.
This is a positive quantity that receives large contributions when both modes are considerably large at some locations. Moreover, for two identical modes it becomes their inverse
participation ratio (IPR). The latter is a well-known measure for the localization, taking
on a value of $N$ for modes localized on a single point (on the lattice) and
$1$ for constant modes.  In fact, we find $I\simeq 1$ also for two normalized
modes with independent Gaussian distributed amplitudes at each site.  Only
modes that are similar {\it and} localized generate large values of $I$.

\enlargethispage{\baselineskip}

\begin{figure}[t]
\includegraphics[width=0.9\linewidth,clip=true,trim=170 \value{scutbottom} \value{scutright} \value{scuttop}]{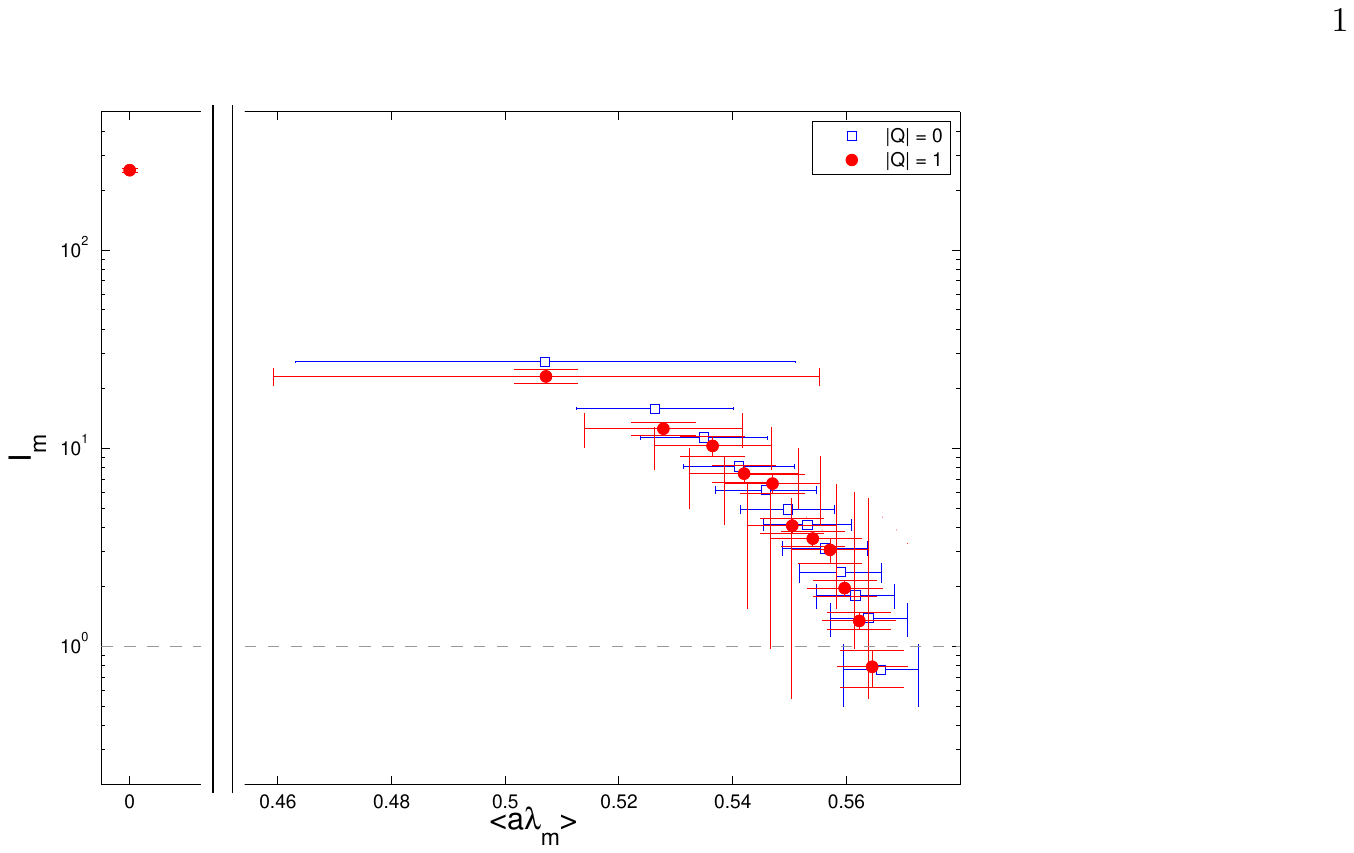}
\caption{The interlocalization $I$, \eqn(\ref{eqn interlocalisation}), for
  matched modes (see text) as a function of the averaged overlap eigenvalue
  $\langle a\lambda_m\rangle$ (in lattice units), on a logarithmic scale and ensemble averaged.  The
  maximal possible value for $I$ on our lattice is $24^3\cdot 4\simeq 5.5\cdot 10^4$,
  whereas delocalized modes yield $I\simeq 1$, indicated by the dashed gray
  line. Horizontal error bars visualize the spreads of the eigenvalues.}
\label{fig matched modes interlocalisation}
\end{figure}

We utilize $I$ for matching the overlap and staggered modes\footnote{In \cite{Hoellwieser:2009} the positions of the highest peaks were used to reveal similarities between overlap and staggered modes.}. We start by
taking the lowest overlap mode and pair it with the staggered mode that has
the largest interlocalization with it. Going up in the overlap spectrum we
continue this matching procedure in the same way, but use only those staggered
modes that have not yet been paired up with a lower overlap mode. In
\fig\ref{fig matched modes interlocalisation} we plot the interlocalization
values for the lowest modes matched in this way as a function of the
corresponding overlap eigenvalue $\lambda_m$. From this plot it is clear that
the zero modes (of the overlap operator, near-zero modes of the staggered
operator) are matching close to perfectly. The value of 300 is actually in the same order of magnitude as the IPR of the
individual overlap zero modes\footnote{The associated low-lying staggered
  modes have a slightly larger IPR, around 400, presumably because in contrast
  to the overlap operator the staggered operator is ultralocal.}.  
Then $I$ drops quickly and after a few modes it reaches the
reference value $1$ discussed above.

\section{Polyakov loops as defects/traps}
\label{sect polyakov}

\begin{figure}[b]
\includegraphics[width=0.9\linewidth]{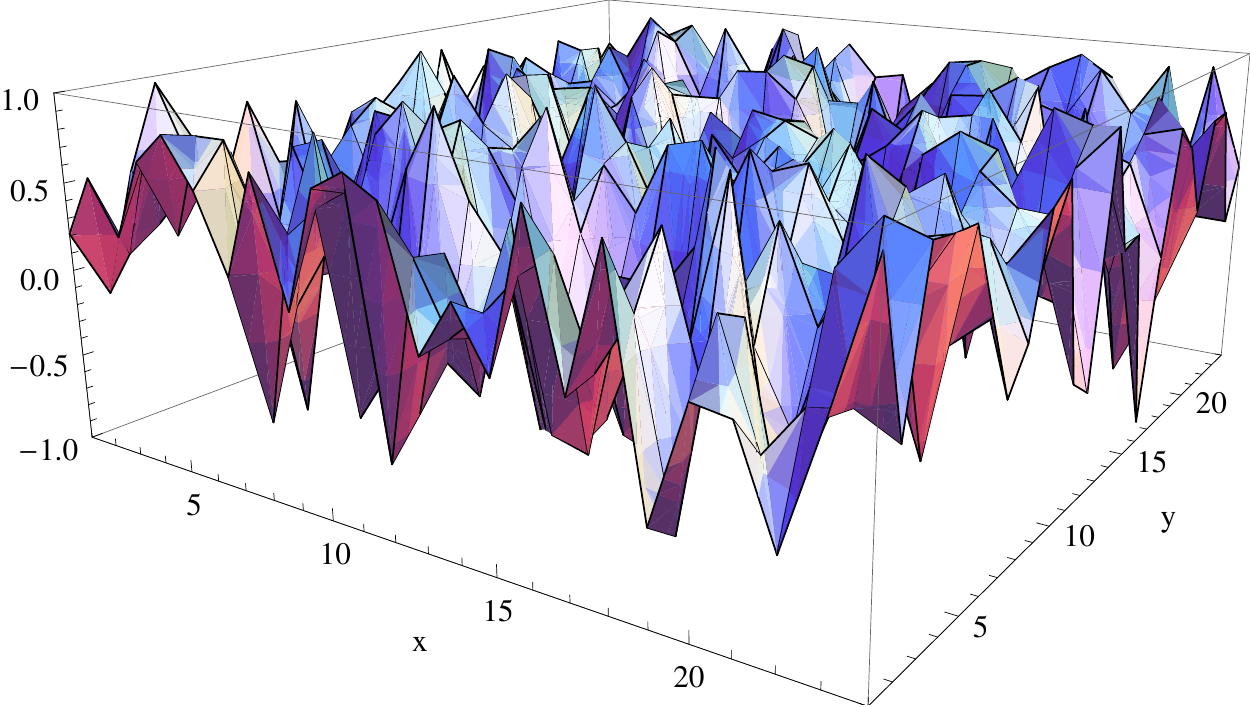}\\
\includegraphics[width=0.9\linewidth]{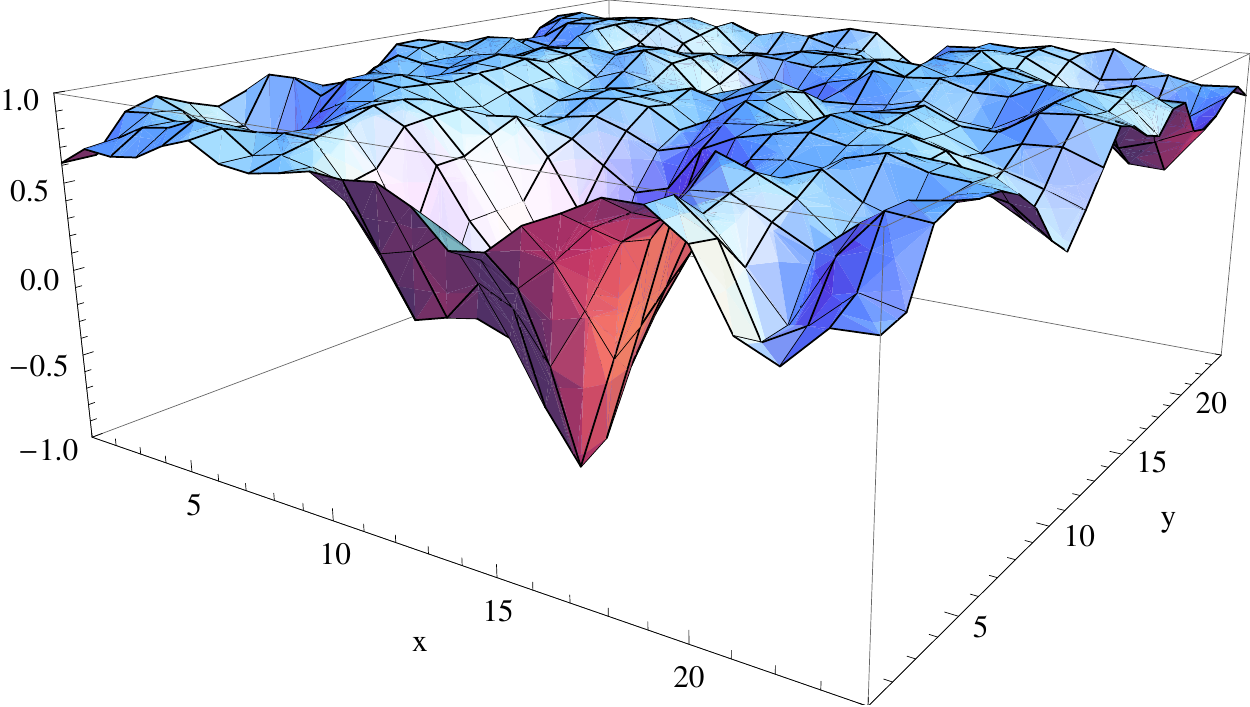}
\caption{Profile of the unsmeared (top) and smeared (bottom) Polyakov loop for the same configuration and in
  the same plane as in Fig.~\protect\ref{fig profile modes}.  By the naked eye nothing seems
  particular in this plane for the unsmeared case, whereas the smeared
  Polyakov loop actually takes its minimum (-0.68, compared to an average
  smeared Polyakov loop of 0.80) at the hotspot visible in the fermion modes
  in Fig.~\protect\ref{fig profile modes}. (A correlation of the unsmeared Polyakov loop to fermion modes is exposed by virtue of statistical measurements, see text.)}
\label{fig profile polloops}
\end{figure}

As an appetizer of our main finding we show in \fig\ref{fig profile polloops}
the Polyakov loops (of one example configuration) in the lattice plane, where
the lowest overlap and staggered eigenmodes take on their maximum, as shown
in \fig\ref{fig profile modes}. The Polyakov loop is dominated by UV
fluctuations at the scale of the lattice spacing as almost every lattice
observable. Therefore it is virtually impossible to see any structures in it.

We applied 6 sweeps of APE smearing \cite{Albanese:1987,Falcioni:1985} with
$\alpha=0.55$ to the configuration, which leads to a smoother Polyakov loop
landscape. Indeed, an island of Polyakov loop with opposite sign emerges at
the location of the maximum of the lowest Dirac mode on the original unsmeared
configuration. A similar profile becomes visible after simply averaging the
(traced) Polyakov loops with their neighbors.

Let us stress, that the lower panel of \fig\ref{fig profile polloops} is the
only occasion that we present a smeared result. We now return to correlation
functions of unsmeared Polyakov loops for the rest of this paper.

\begin{figure}[!b]
\includegraphics[width=\linewidth,clip=true,trim=170 \value{scutbottom} \value{scutright} \value{scuttop}]{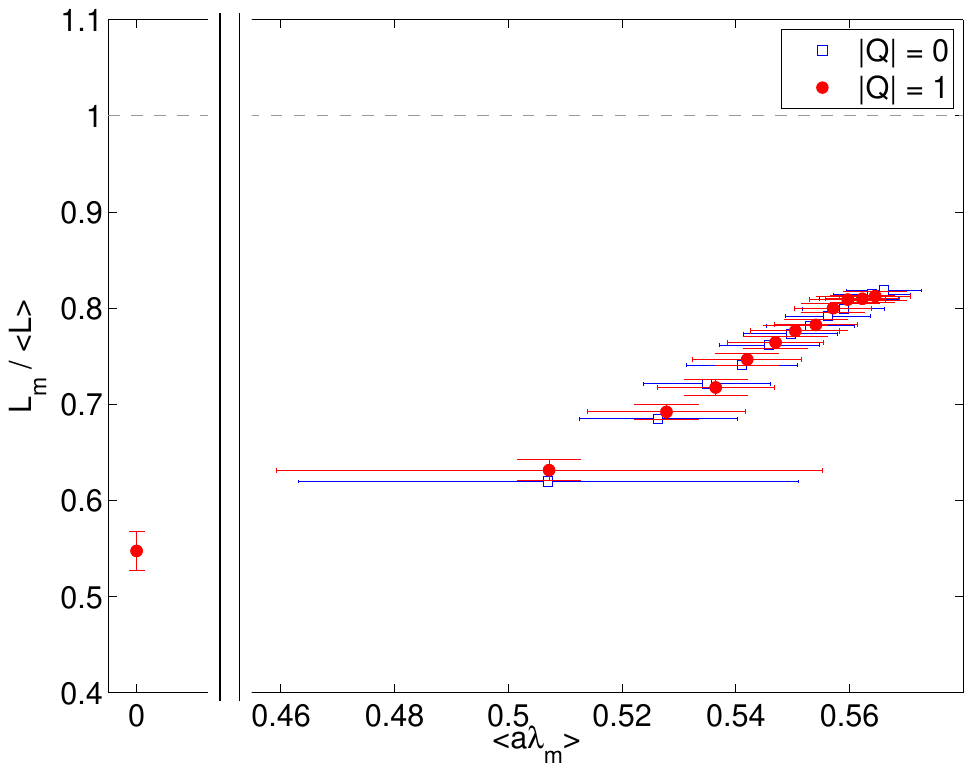}

\includegraphics[width=\linewidth,clip=true,trim=170 \value{scutbottom} \value{scutright} \value{scuttop}]{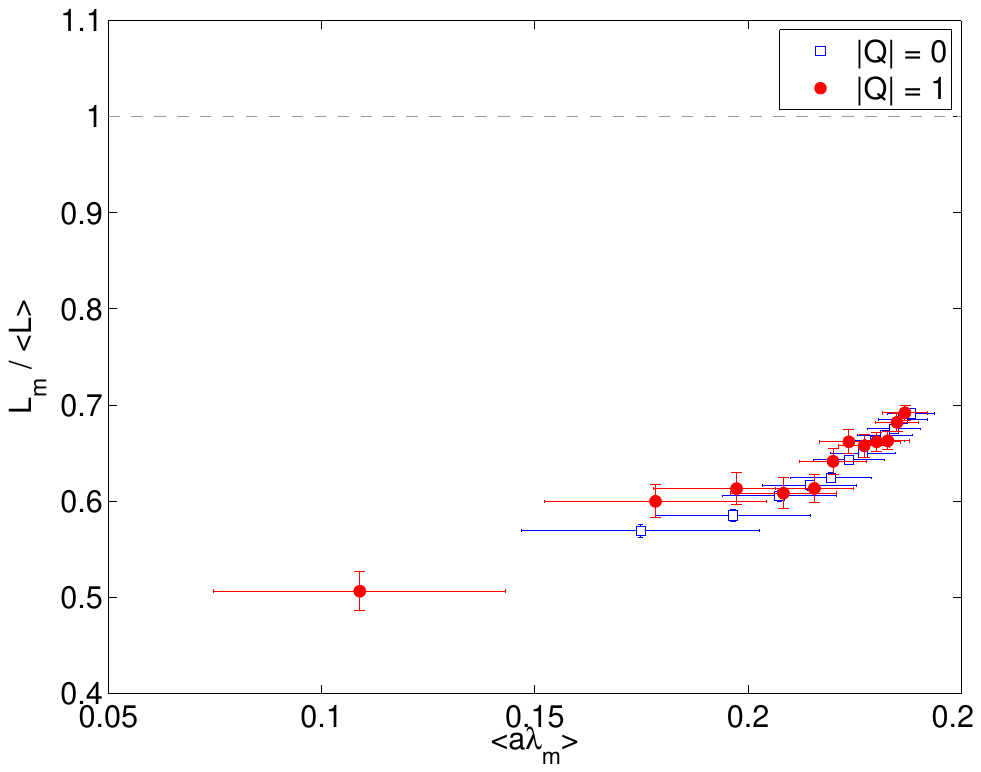}
\caption{The ratio of ``Polyakov loops as averaged by low-lying modes'' $L_m$,
  \eqn(\protect\ref{eqn loops seenby modes}), to the average Polyakov loop
  for overlap (top) and staggered modes (bottom)
  as a function of the corresponding averaged eigenvalue
  $\langle a \lambda_m\rangle$. Horizontal error bars visualize the spreads of the eigenvalues.}
\label{fig polloop modes one}
\end{figure}

To check the correlation of Polyakov loop islands and low Dirac modes in
  a quantitative way, we define ``Polyakov loops as averaged by a
particular mode'', i.e. Polyakov loops weighted with the density of a
  normalized Dirac mode, cf.\ \cite{Bruckmann:2005c},
\begin{equation}
 L_m := \sum_x |\psi_m(x)|^2 L(x)\,.
\label{eqn loops seenby modes}
\end{equation}
This quantity is restricted to the interval $[-1,1]$, just like the
Polyakov loop $L$.  It is clear that a wave-like mode $\psi_m$ with
approximately constant amplitude yields an $L_m$ close to the average Polyakov
loop $\sum_x L(x)/V$, whereas a strongly localized mode picks the Polyakov
loop in that region. If the latter happens to be an island of ``wrong''
Polyakov loops, $L_m$ tends to zero or even becomes negative.

\Fig\ref{fig polloop modes one} shows the ratio of $L_m$ averaged over
different configurations and the average Polyakov loop, both for the low-lying
overlap and low-lying staggered modes. The Polyakov loops averaged by the
low-lying modes are indeed smaller than on average. Higher modes, on the
contrary, tend to see the average Polyakov loop.

\begin{figure}[!]
\includegraphics[height=1.1\twopicwidth]{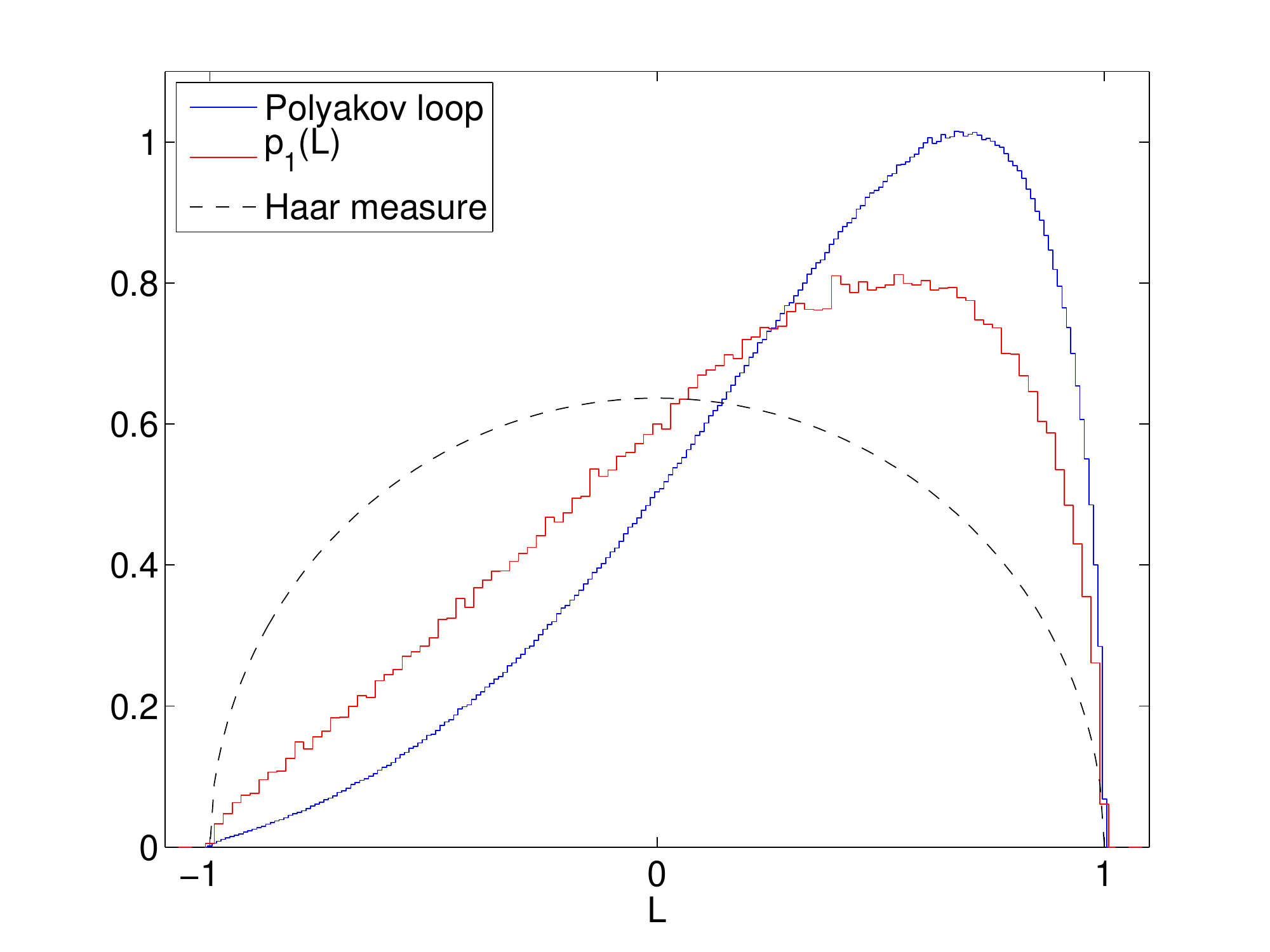}
\caption{The ``Polyakov loop distribution as seen by the lowest overlap mode'', $p_1(L)$ from
  \eqn(\protect\ref{eqn new probability}),  in 
  the $Q=0$
  sector 
  compared to the Polyakov loop
  distribution and the Haar measure (valid in the low temperature phase).
   }
\label{fig polloop modes two}
\end{figure}

The connection between local Polyakov loops and low-lying Dirac modes
is confirmed from a slightly different perspective by the following
``Polyakov loop distribution as seen by a mode''. For that we weight the probability of Polyakov loops $L$ by the amplitudes
of the low-lying modes at those positions $x$ where $L(x)=L$,
\begin{equation}
 p_m(L)=\sum_x\delta_{\epsilon}(L-L(x))|\psi_m(x)|^2\,,
\label{eqn new probability}
\end{equation}
in continuous notion with some smeared delta-function $\delta_{\epsilon}$ (in practice we rescale histograms). The quantity $L_m$ is recovered by the $L$-expectation value with this probability,
\begin{equation}
 L_m=\int dL\, p_m(L)\, L\,,\qquad \big(\int dL\, p_m(L)=1\big)\,. 
\end{equation}
Thus the probability $p_m$ visualizes how the global quantity $L_m$ is
generated by modifying the distributions of the local Polyakov loops.

In \fig\ref{fig polloop modes two} we show this probability for the lowest modes of the overlap operator, $p_1(L)$, on $Q=0$ configurations. One can clearly see that low fermion modes enhance low Polyakov loops down to $L\simeq -1$.

\section{Interpretations}
\label{sect interpretation}

\subsection{Effective Matsubara frequencies}

In a background of constant temporal and vanishing spatial gauge fields
the influence of the Polyakov loop on the Dirac spectra is very clear. We
first diagonalize the Polyakov loop introducing its phase
$\varphi$,
\begin{eqnarray}
 \prod_{x_0=1}^{N_t}U_0(x_0) & = & \exp\big(i\varphi\left(\begin{array}{cc} 1 & \\ & -1\end{array}\right)\big)\,,
\label{eqn polloop phase first}\\
L & = & \cos\varphi\qquad \varphi \in [0,\pi]\,.
\label{eqn polloop phase second}
\end{eqnarray}
We gauge it into the last time slice, where it effectively changes the
temporal boundary condition.

The lowest modes of the free Dirac operator in the presence of this Polyakov loop
 are constant in space and
plane waves in time, $\exp(i p x_0 T)$. The 
quantum numbers 
$p$ are governed by a
combination of the antiperiodic temporal boundary condition for fermions and
the Polyakov loop phase, namely $p=\pi\pm\varphi+2\pi\mathbb{Z}$, where the
different signs emerge from the different color components, see 
\eqn(\ref{eqn polloop phase first}). 

The eigenvalues of the free Dirac operator are these 
numbers
multiplied by the
temperature. The lowest ones,
\begin{equation}
 \lambda_{\rm M}^{\mbox{cont.}}=(\pi-\varphi)T\,,
\label{eqn eff MF}
\end{equation}
we name {\it effective Matsubara frequencies}. These hold in the limit $N_t \to\infty$, whereas on the lattice one has
\begin{equation}
 \lambda_{\rm M} = \frac1a \sin\left(\frac{\pi-\varphi}{N_t}\right)\,.
\label{eqn eff MF latt}
\end{equation}

At high temperatures the Polyakov loop (at fixed lattice spacing) becomes
trivial, $L\to 1$, hence $\varphi\to 0$ and the Matsubara frequency is
$\lambda_{\rm M}=\frac1a \sin\left(\pi/N_t\right)$. A more realistic estimate
is obtained by using the average Polyakov loop at our temperature $\langle L
\rangle =0.37$, from which we obtain the effective Matsubara frequency in
lattice units ($T=1/N_t a$) as
\begin{equation}
 \lambda_{{\rm M}} a = \sin\left(\frac{\pi-\arccos 0.37}{4}\right)=0.47\,,
\end{equation}
which is the same order of magnitude as the lower end of the bulk of eigenvalues we measured (consistent with the findings of \cite{Gavai:2008}). 

The main point of these considerations is that ``wrong'' Polyakov loops, $L=-1$ with $\varphi=\pi$, would lead to a vanishing effective Matsubara frequency, $\lambda_{\rm M}=0$, and thus to the lowest Dirac eigenvalues. 

Of course, in realistic configurations one has to take into account that the Polyakov loop varies in space and that nontrivial spatial links are present.
Both will change the Dirac eigenvalues away from the free ones. Nonetheless, the tendency that ``wrong'' Polyakov loops give rise to smaller eigenvalues
persists and explains our finding about their pinning nature.

\subsection{Topological objects}

Topological excitations of the gauge field and their zero modes are an
attractive hypothesis to explain low-lying Dirac modes in Yang-Mills theory
(and QCD). The chiral condensate at zero temperature is thought
of as due to instantons of realistic size and density and the first conjecture about the metal-insulator transition at finite temperature was based on instanton ensembles \cite{Diakonov:1984vw,GarciaGarcia:2005vj}.

The natural topological excitations at finite temperature are magnetic
monopoles.  They appear as self\-dual or antiselfdual solutions of the
Yang-Mills equations of motion at finite temperature. As they are also
electrically charged, they are called dyons. Dyons can be constituents of
calorons (finite temperature instantons) \cite{Kraan:1998,Lee:1998}, but may
exist in isolation as well \cite{Gross:1981}.

In gauge group $SU(2)$, there are two dyons (and two antidyons) with opposite
magnetic charge. One sort of these dyons is characterized by properties that
exactly match our findings: the Polyakov loop at their core is $-1$ and they
possess zero modes with the physical antiperiodic boundary conditions
\cite{Callias:1978,Nye:2000} (just like the constant configurations discussed
above). The other sort has Polyakov loop $+1$\footnote{Monopoles also
    appear as defects \protect\cite{Hooft:1981} of the Polyakov gauge
    \protect\cite{Weiss:1981}, where they have $L=-1$ or $L=+1$ by
    definition.} and zero modes with periodic boundary conditions.

\begin{figure}[!t]
\includegraphics[width=0.9\linewidth]{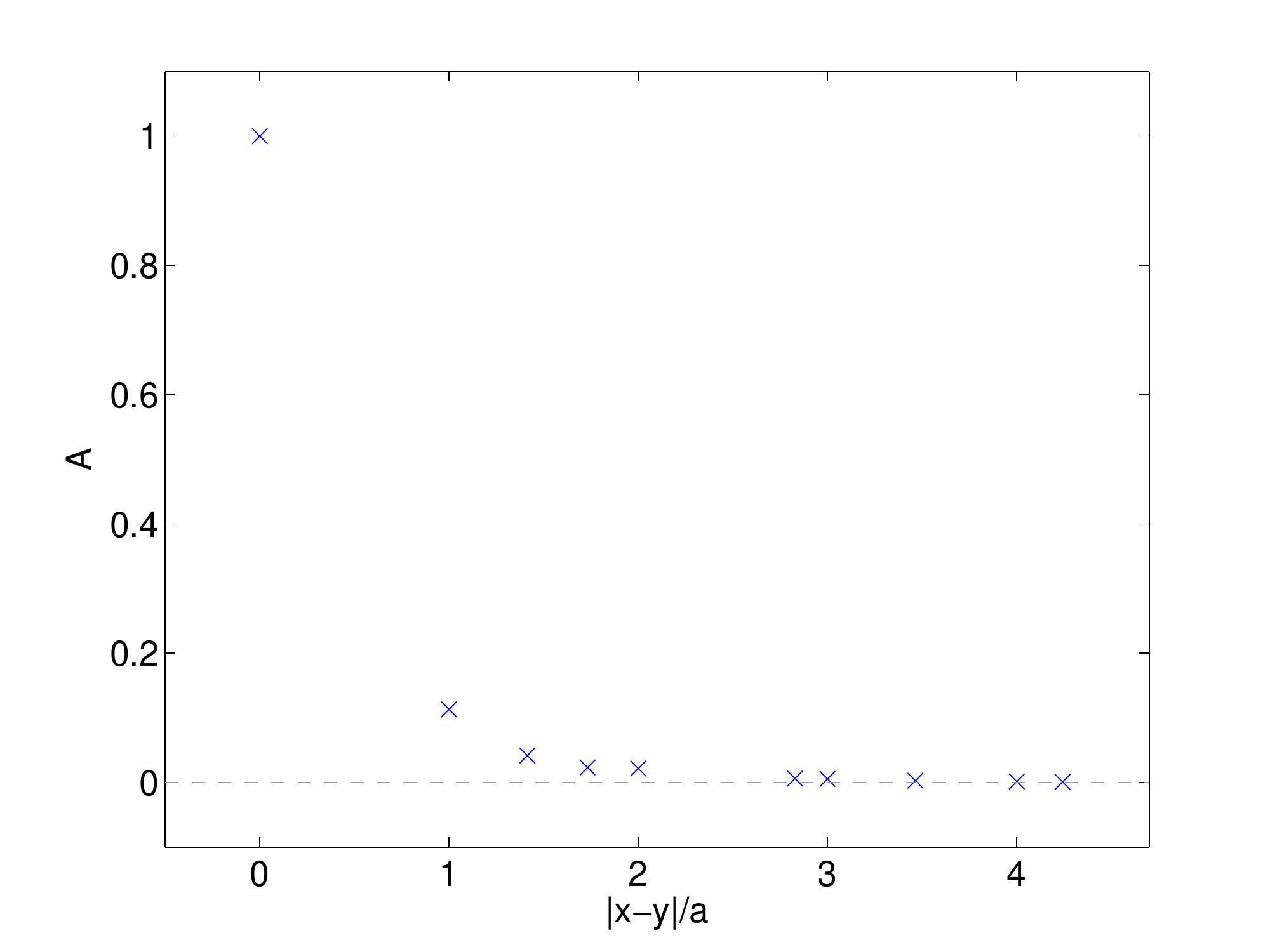}\\
\includegraphics[width=0.9\linewidth]{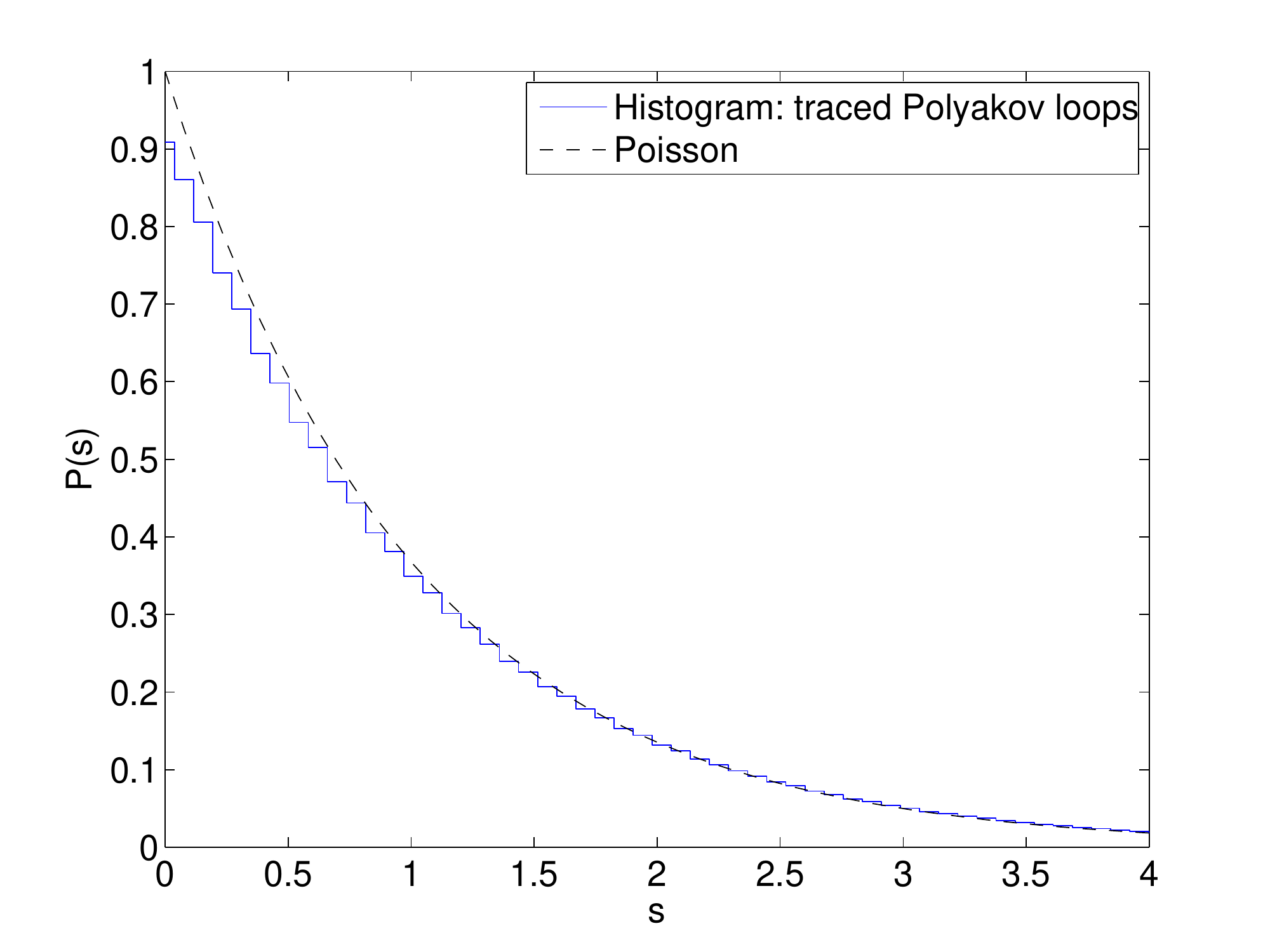}
\caption{Independence of local Polyakov loops. Top: the auto-correlation 
$A:=\left(\langle L(x)L(y)\rangle - \langle L(x)\rangle\langle L(y)\rangle \right)/$ $\left(\langle L^2(x)\rangle - \langle L(x)\rangle^2 \right)$
as a function of the lattice distance $|x-y|/a$ 
(containing the same information as the free energy). 
Bottom: the level spacing distribution (of the Polyakov loop trace $L$) compared to the Poisson distribution.}
\label{fig polloop properties}
\end{figure}

In dilute ensembles of dyons most of the zero modes should remain low-lying
modes. Then the first sort of dyons could explain the
localized modes we analyzed, while the second sort could
be responsible for low-lying periodic modes. At a positive average Polyakov
loop one expects fluctuations to $-1$ much less frequent than those to $+1$,
cf.\ \fig\ref{fig polloop properties}. Hence the first sort of dyons
could give independent low-lying modes, while the second sort
could yield a condensate at periodic boundary conditions. This different
appearance can be made quantitative by the different fractions of unit
topological charge the two dyon sorts have, which are such that the $-1$
dyons are indeed heavier as they have a larger (classical) action, see
\cite{Bornyakov:2008im} for lattice evidence of this picture and 
\cite{Bruckmann:2008d} for a simple model.

From the relation of magnetic monopoles to both low-lying modes and
topological charge a crucial test of this picture is to compare these two
quantities at given temperature and volume.  Above the finite temperature
transition fluctuations of the topological charge decrease sharply. This can
be clearly seen by looking at how the topological susceptibility decreases at
higher temperatures. As a result, at high temperature,
topological objects presumably form a dilute gas of non-interacting
objects. From the index of the overlap Dirac operator we have full information
about the fluctuations of the {\it total topological charge}. Assuming that in
the dilute gas topological objects are uncorrelated, this can be used to
compute the density of topological objects that in turn can be compared to the
density of localized Poissonian Dirac eigenmodes.

In Table \ref{tab:charge_distribution} we show the probability of different
topological sectors extracted from the index of the overlap Dirac
operator. From the low occurrence of the topological charge sector $\pm 1$ and
in particular of the sector $\pm 2$ it is indeed clear that topological
objects form a very dilute gas. In these volumes it very rarely happens that
there is more than one topological object on any given configuration. Since we
need only an order of magnitude estimate of the density of topological objects
we will ignore the probability of two or more objects occurring on any single
configuration. In this approximation we can shortcut the calculation of the
density of topological objects, and the average number of topological
objects per configuration is just given by the probability of the $|Q|=1$
sector. On our volumes it is between $0.04$ and $0.1$ which is clearly far too
small to account for the few localized Poissonian modes we found on average
per configuration.

This comparison rules out models based on uncorrelated gluonic objects that carry both (O(1)) topological charge and (antiperiodic) zero modes. However, it does not exclude combinations of topological objects in which the topological charge cancels, like 
instanton-antiinstanton molecules 
originally suggested to be present with light dynamical quarks \cite{Ilgenfritz:1994nt} or molecules of dyons called 'bions' carrying one near zero mode \cite{Unsal:2007jx}.

\begin{table}
\begin{tabular}{|c|c|c||c|c|c|}
    \hline
  $|Q|$ & 0 & 1 & 0 & 1 & 2 \\
      \hline
  ${\cal P}_Q$ & 0.958(6)  & 0.042(6) & 0.89(1) & 0.105(9) & 0.009(3) \\
    \hline
\end{tabular}
  \caption{The probability of different charge sectors in the $16^3\times 4$
    (first two columns) and $24^3\times 4$ ensemble (last three columns). The
    probabilities of charge sectors with the same magnitude but opposite sign
    have been added. 
       \label{tab:charge_distribution}}
\end{table}

\section{Random matrix model using the staggered Dirac operator}
\label{sect rmt}

\begin{figure*}
\includegraphics[height=1.1\threepicwidth]{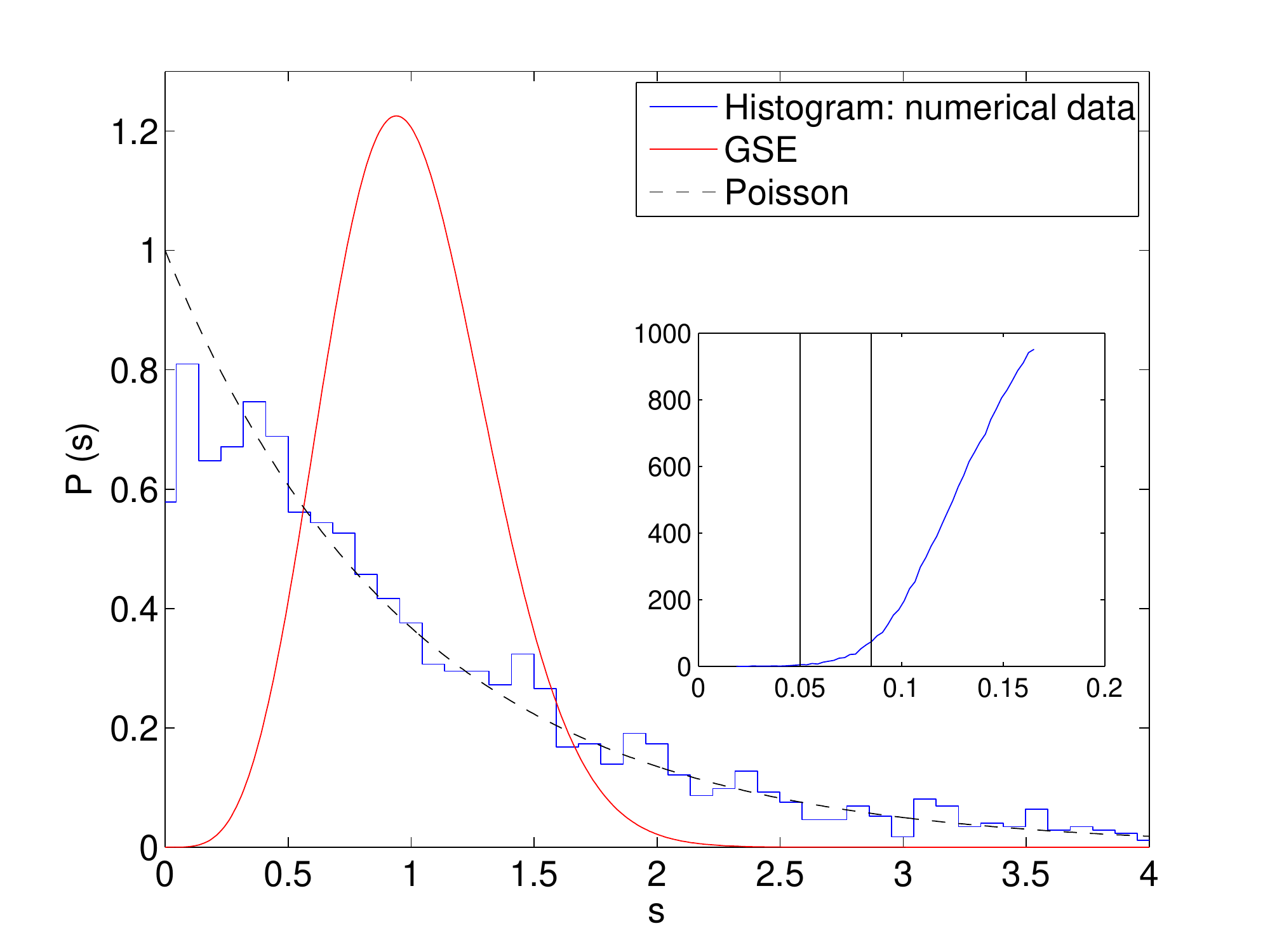}\hspace*{-0.9cm}
\includegraphics[height=1.1\threepicwidth]{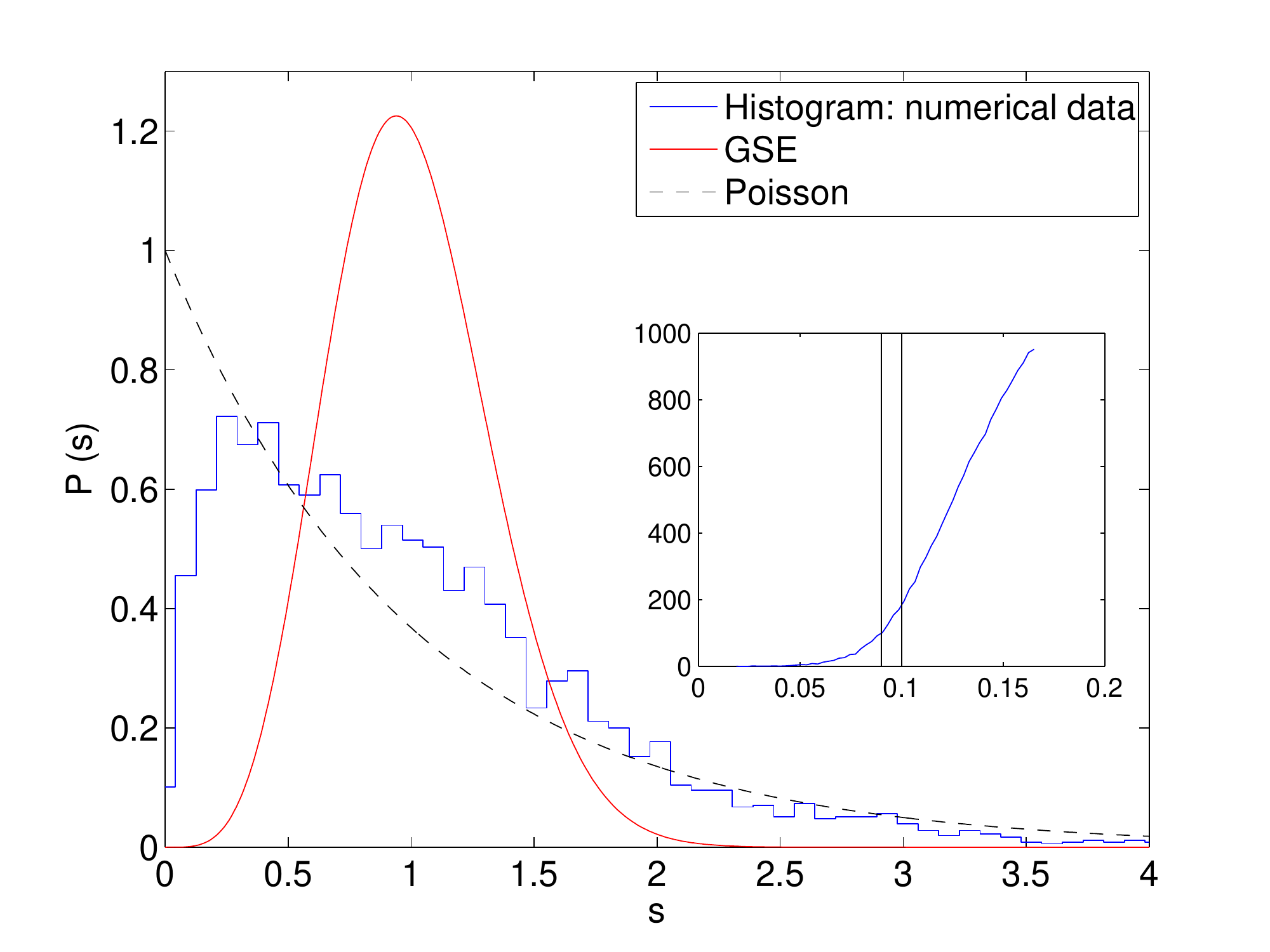}\hspace*{-0.9cm}
\includegraphics[height=1.1\threepicwidth]{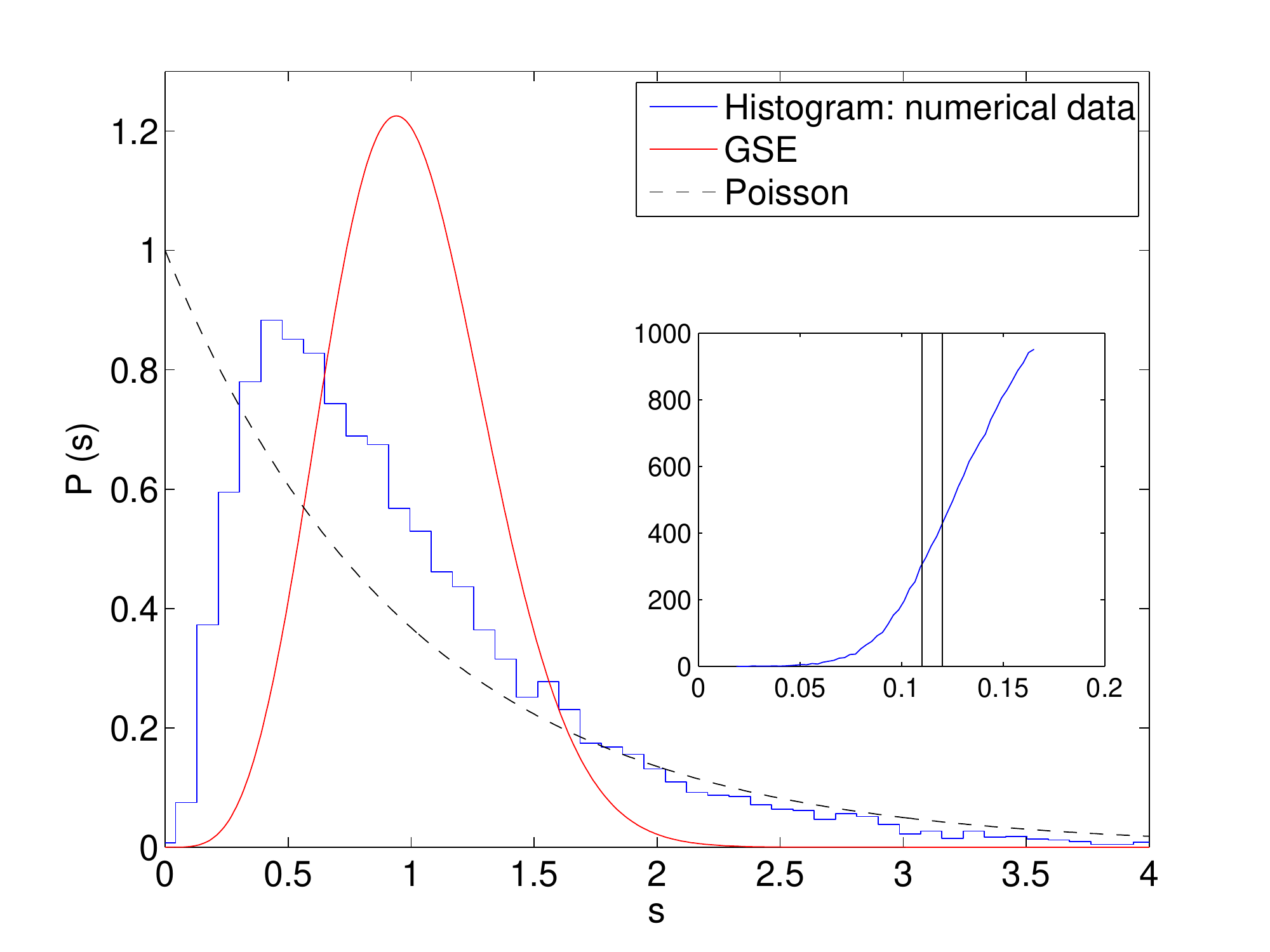}

\includegraphics[height=1.1\threepicwidth]{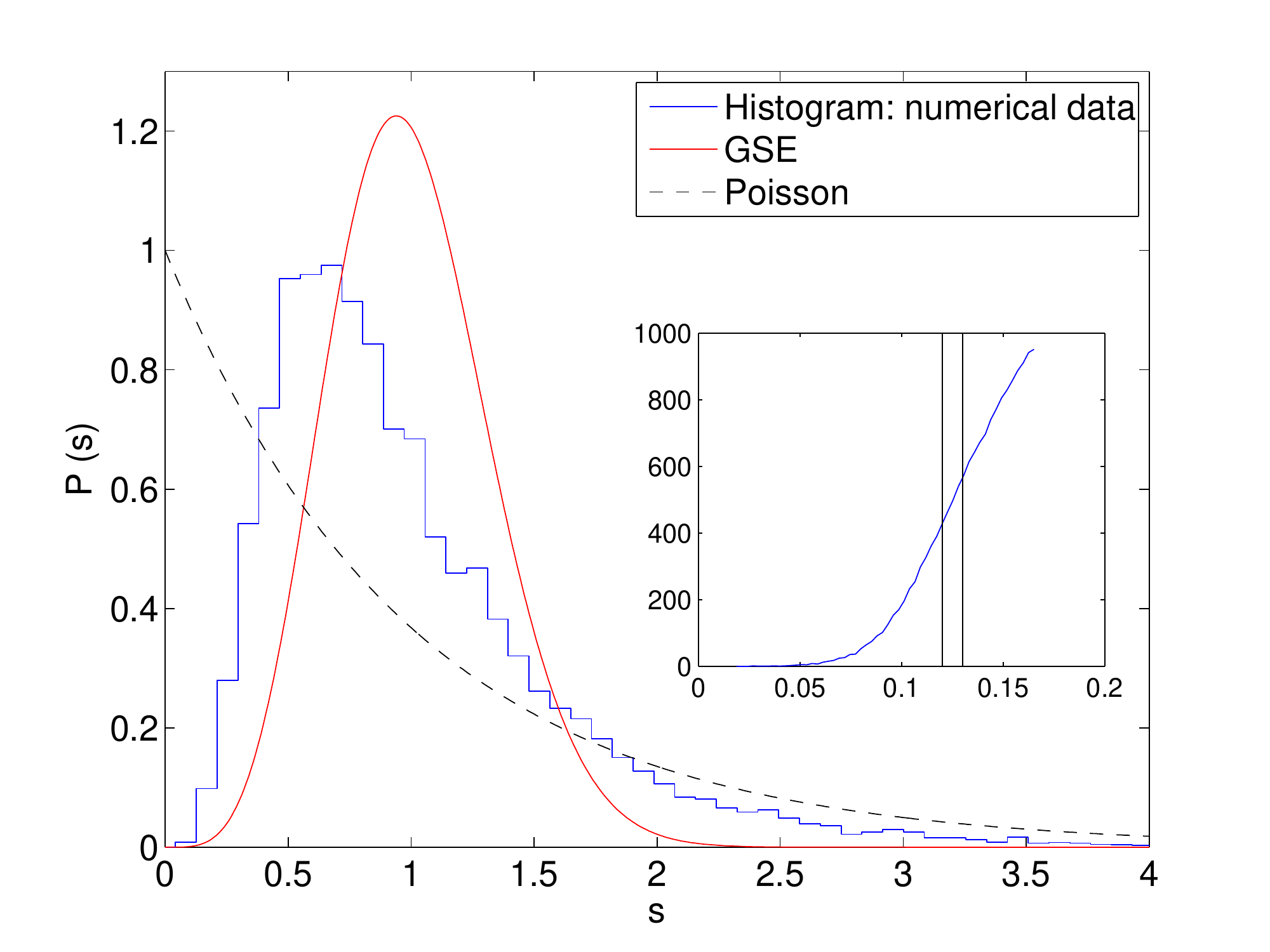}\hspace*{-0.9cm}
\includegraphics[height=1.1\threepicwidth]{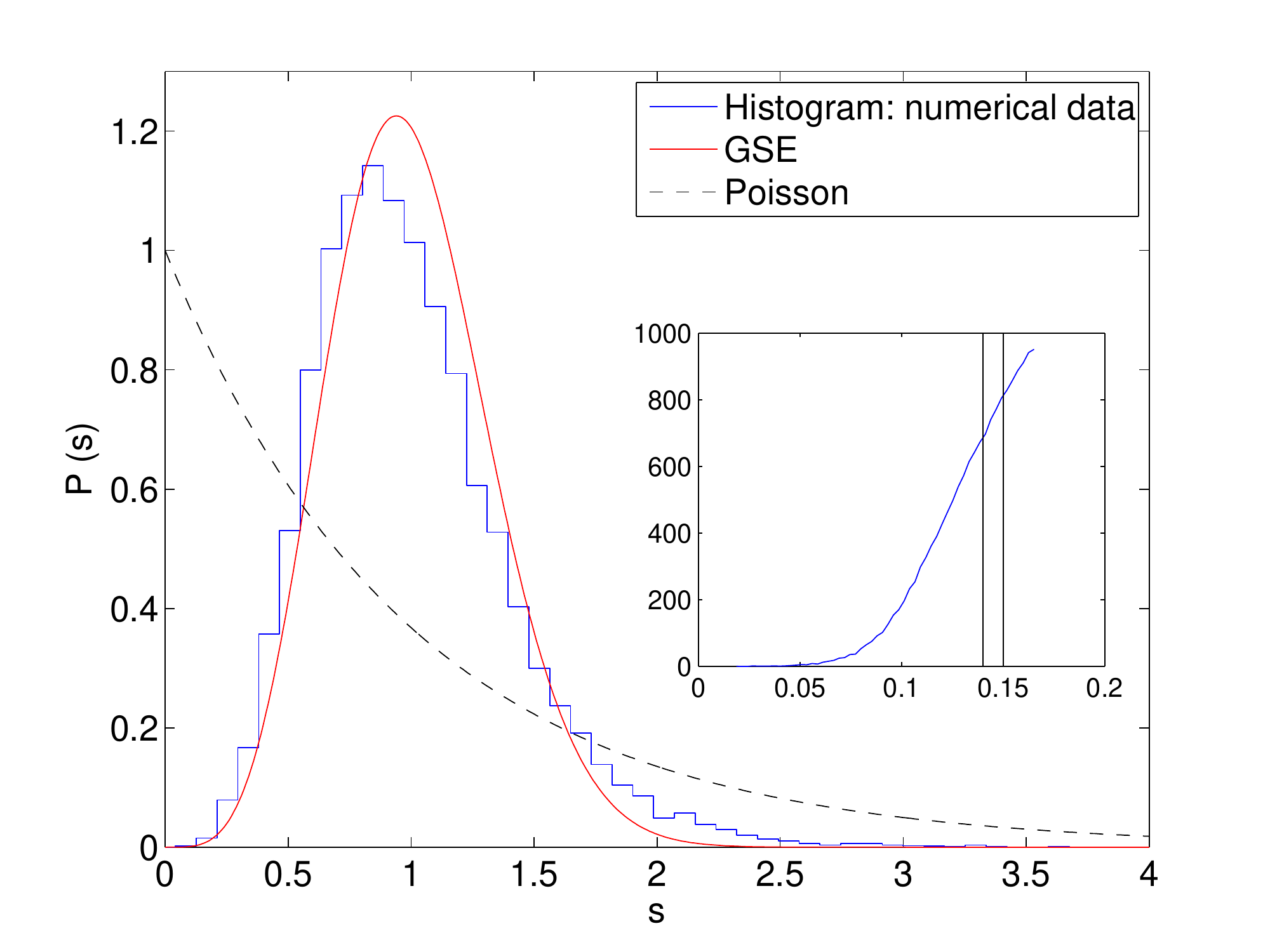}\hspace*{-0.9cm}
\includegraphics[height=1.1\threepicwidth]{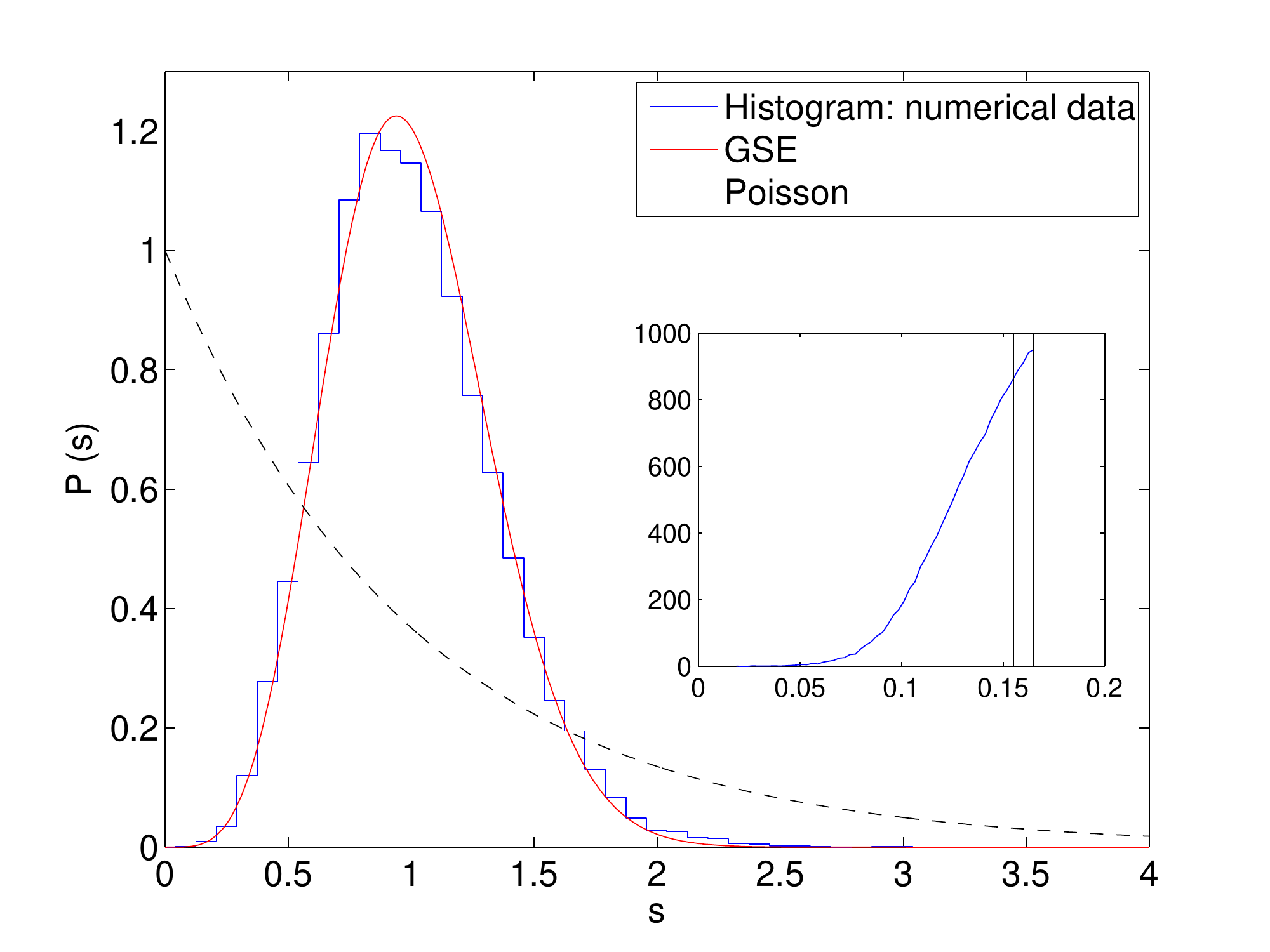}

\caption{Spacing distributions in several parts of our RMT spectrum plotted along with
the GSE prediction and the Poissonian distribution. The parameters were fixed as discussed in the text.
The data was obtained by an ensemble average over $2000$ random matrices.}
\label{sparsermtspacfigure}
\end{figure*}

Transitions between correlated and uncorrelated eigenvalues like the one observed in
the lattice data can be described by RMT models in various ways. One of the
simplest possible ans\"atze is to start with a diagonal matrix, whose entries are
uncorrelated random numbers, and add a matrix taken from one of the Gaussian
ensembles. This model 
shows the desired transition in between parts of its spectrum with different
eigenvalue density,
however, the interpolating spacing distributions 
are different
from the ones in the spectrum of the staggered operator \cite{Schierenberg:x}.  
A possible explanation is the
sparseness of this operator and the spatial information that is contained in
the next-neighbor interactions. In contrast, the full matrices from
the Gaussian ensembles blindly connect all diagonal elements with equal
strength.

\subsection{Motivation}

We construct a better suited random matrix model, based on sparse matrices, in the following. This model can be nicely motivated by our previous
findings. Concerning the Polyakov loop, which will be connected to a random potential, three properties are relevant:
(i) the distribution of local Polyakov loops extends to negative values and thus small effective Matsubara frequencies, cf. \fig\ref{fig polloop modes two},
(ii) Polyakov loops become independent quickly with their distance and therefore
(iii) the level spacings of the Polyakov loop trace $L$ -- neglecting the spatial information -- are distributed according to a Poissonian distribution.
The last two properties are depicted in \fig\ref{fig polloop properties}.

Hence the Polyakov loops could provide the Poissonian ingredient for the statistics of the Dirac eigenvalues. To become more concrete, we
remind the reader of the definition of the staggered Dirac operator,
\begin{equation}
D_{xx'} = \frac1{2a} \sum_{\mu=1}^4 \eta_\mu(x)\left[\delta_{x+\hat\mu, x'}U_\mu(x) - \delta_{x-\hat\mu, x'}U_\mu^\dagger(x')\right]\,,
\end{equation}
with $\eta_\mu(x) = \left(-1\right)^{\sum_{\nu<\mu}x_\nu}$ and $U\in SU(2)$.  We split this operator in the temporal and spatial part
\begin{equation}
D = D^{TE} + D^{SP}\,.
\end{equation}
$D^{TE}$ contains the hopping terms in the temporal direction, whereas spatial hoppings are included in $D^{SP}$.

\begin{figure}
\includegraphics[width = 0.9\linewidth]{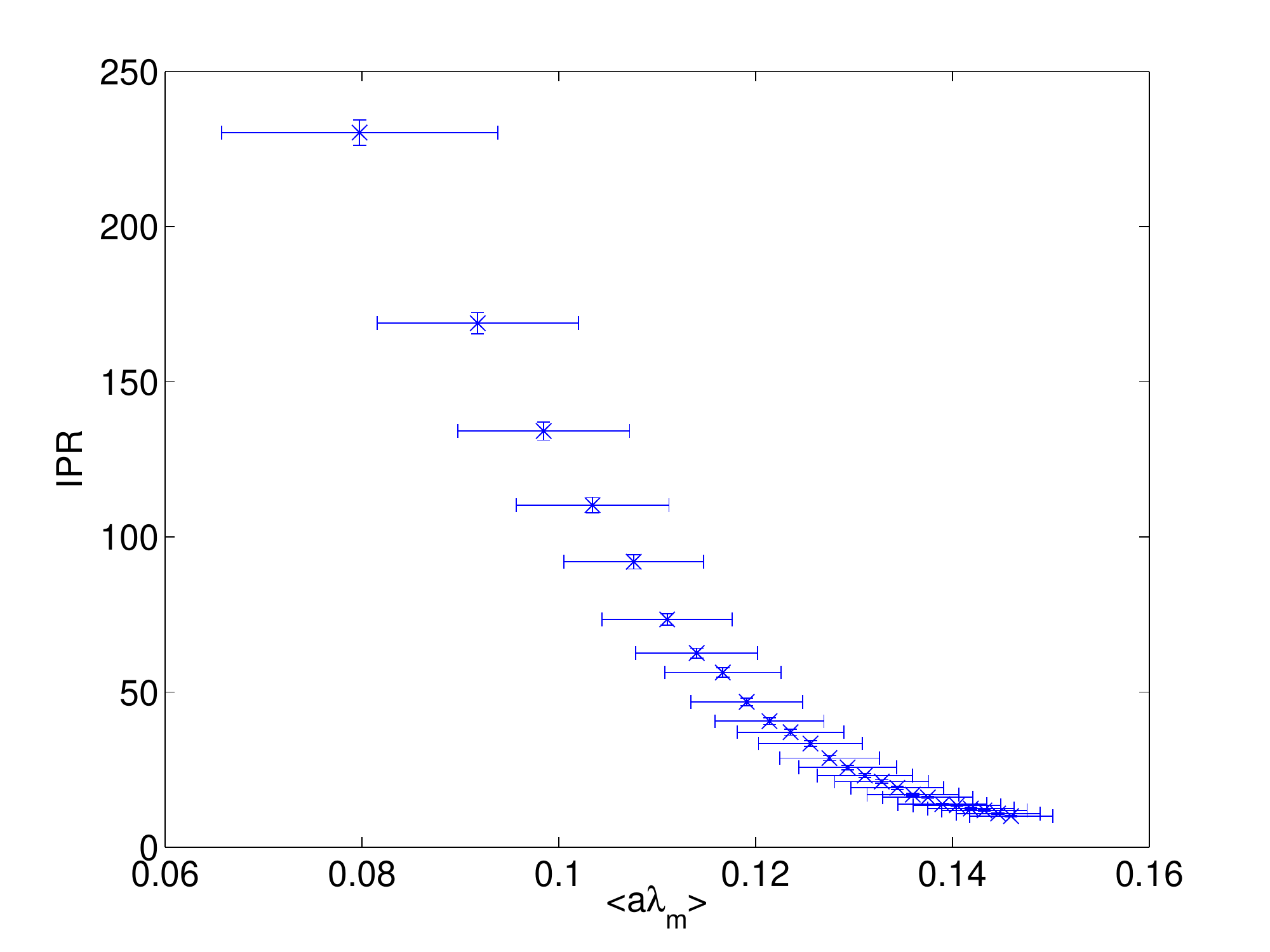}
\includegraphics[width = 0.9\linewidth]{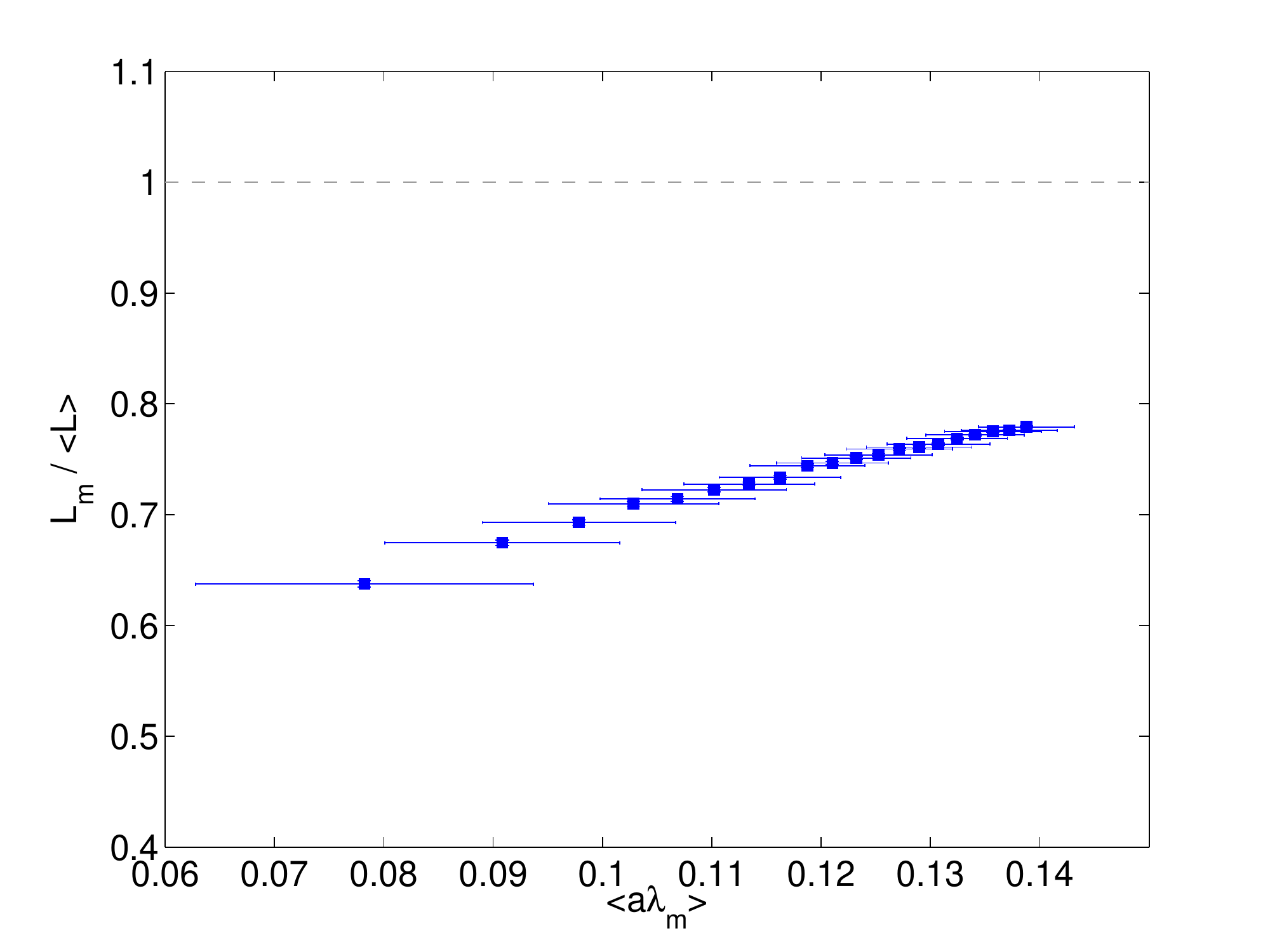}
\caption{Top: Inverse participation ratios of the 25 lowest-lying eigenmodes for the RMT model, plotted versus the average eigenvalue of
that mode. The data was obtained by an ensemble average over $2000$ random matrices. 
Bottom: diagonal entries ``as averaged by the lowest modes'' of the RMT model, $\vartheta_m$ from \eqn(\protect\ref{eqn theta averaged}), divided by the average diagonal entry, to be compared with \fig\protect\ref{fig polloop modes one}.
Horizontal error bars visualize the spreads of the eigenvalues.}
\label{sparsermtlocfigure}
\end{figure}

The temporal part has a block-diagonal structure, with one block for each
spatial site. We want to approximate each block and thus the whole Dirac operator 
by restricting it to the subspace of the smallest
eigenvalue quadruplet of the temporal operator at each spatial lattice
site. The quadruplet consists of two eigenvalue pairs with opposite sign. The
plus-minus degeneracy is necessary to conserve chiral symmetry, whereas the
exact two-fold degeneracy\footnote{In RMT language, this is Kramers degeneracy
  of the GSE.} has to be kept in order to have the right RMT universality
class we find in the SU(2) staggered spectra.

In this basis, the temporal part of the Dirac operator can be brought in the form
\begin{equation}
D^{TE(n=0)}_{\vec x \vec x} = \begin{pmatrix} -\theta_{\vec x} & 0 & 0 & 0 \\ 0 & -\theta_{\vec x} & 0 & 0 \\
 0 & 0 & \theta_{\vec x} & 0 \\ 0 & 0 & 0 & \theta_{\vec x}\end{pmatrix}\,,
\end{equation}
with
\begin{equation}
\theta_{\vec x} = \frac1a \sin\left(\frac{\pi-\varphi_{\vec x}}{N_t}\right)\,,
\end{equation}
which resemble the effective Matsubara frequencies on the lattice, as given in \eqn(\ref{eqn eff MF latt}), however as a function of the
local rather than averaged Polyakov loop phase $\varphi_{\vec x}$.  Therefore, $D^{TE(n=0)}$ is a diagonal matrix with very weakly coupled entries, as argued above.
Physically speaking, the temporal part represents a (chiral) random potential.

The spatial part becomes in the restricted basis
\begin{equation}\label{sp0dirac}
D^{SP(n=0)}_{\vec x, \vec x + \ihat } = \begin{pmatrix} U_{\vec x, \vec x + \ihat } & V_{\vec x, \vec x + \ihat } \\
V_{\vec x, \vec x + \ihat } & U_{\vec x, \vec x + \ihat }\end{pmatrix}\,,
\end{equation}
where the 2-by-2 matrices $U$ and $V$ can be shown to represent real quaternions.

These have a concrete meaning:
$U$ connects eigenvalue pairs of equal sign, and is therefore responsible for the GSE-like level repulsion between nearest neighbors.
$V$ on the other hand generates the gap around zero in the spectrum as it connects eigenvalues of different signs.

\subsection{Explicit construction of the model}

We propose a random matrix model based on matrices
\begin{equation}
M = M^{TE} + M^{SP},
\end{equation}
that consist of two parts. $M^{TE}$ has the same diagonal structure like $D^{TE(n=0)}$, i.e.
\begin{equation}
M^{TE}_{\vec x \vec x} = \begin{pmatrix} -\vartheta_{\vec x} & 0 & 0 & 0 \\ 0 & -\vartheta_{\vec x} & 0 & 0 \\
 0 & 0 & \vartheta_{\vec x} & 0 \\ 0 & 0 & 0 & \vartheta_{\vec x}\end{pmatrix}\,,
\end{equation}
where the diagonal entries 
\begin{equation}
 \vartheta_{\vec x} = t(\pi-\phi_{\vec x})
\end{equation}
are random numbers constructed with an overall scale $t$ and a random angle $\phi_{\vec x}\in[0,\pi]$. These quantities are equivalent to the effective Matsubara frequencies $\theta_{\vec x}$ in the continuum, \eqn(\ref{eqn eff MF}), the temperature $T$ and the angle $\varphi_{\vec x}$ of the local Polyakov loop, respectively.

For $\phi_{\vec x}$ we have taken the empirical distribution of the angle $\arccos L(\vec{x})$ of local Polyakov loops, by converting the fine histogram in \fig\ref{fig polloop modes two} accordingly. This yields an asymmetric distribution of $\phi_{\vec x}$ between $0$ and $\pi$ with a maximum below $\pi/2$ (i.e.\ at positive Polyakov loop). Its most important feature, however, seems to be the tail towards the ``trapping'' $\phi_{\vec x}\simeq \pi$ (negative Polyakov loop locally), since we have observed that most of the features of the model persist when using the Haar measure $\sin^2\phi$ as distribution [not shown].

For the spatial part, one first of all needs to fix the periodicity of the underlying space, i.e. an integer $N_s$ such that $\vec x$ is identified with
$\vec x + N_s \ihat$, with unit vectors $\ihat$ in each of the spatial directions. $M^{SP}$, like $D^{SP(n=0)}$, has nonvanishing entries only at positions
that connect next neighbors in that space. Its blocks
\begin{equation}
M^{SP}_{\vec x, \vec x + \ihat } = \begin{pmatrix} u_{\vec x, \vec x + \ihat } & v_{\vec x, \vec x + \ihat } \\
v_{\vec x, \vec x + \ihat } & u_{\vec x, \vec x + \ihat }\end{pmatrix}\,,
\end{equation}
consist of random real quaternions $u$ and $v$, that are RMT counterparts of $U$ and $V$ from \eqn(\ref{sp0dirac}). We have taken them to be Gaussian distributed around zero with their mean deviations $\sigma_{u},\sigma_{v}$ as parameters of the model.

By rescaling all the random matrices $M$ it is clear, that only the ratios of the scales $t$, $\sigma_{u}$ and $\sigma_{v}$ are relevant parameters. To fix them we have measured the ratio of the average determinants\footnote{For real quaternions, the determinant is just the sum over all squared components.} of the empirical quaternionic hopping terms finding $\langle \det V\rangle/ \langle \det U\rangle \approx 1.6^2$ and took this ratio over for the ratio of $\sigma_{u}^2/\sigma_{v}^2$. The remaining ratio $\sigma_u/t=0.2$ ($\sigma_v/t=0.32$)
was put in by hand to obtain desired properties of the RMT model, namely a gap at zero and a
transition between a Poissonian and a GSE spacing distribution.

The eigenvalue density and spacing distribution of this model are plotted in \fig\ref{sparsermtspacfigure} for a spatial extent $N_s = 12$  and $t=1/4$.
We observe a similar transition like in the spectrum of the staggered Dirac operator. Another feature that this model shares with the staggered operator is the
increasing localization of eigenmodes as the corresponding eigenvalues decrease, 
quantified by the IPR's in \fig\ref{sparsermtlocfigure} top.

In order to measure the correlation of the lowest eigenmodes of this RMT model to the diagonal entries, we again define the latter ``as averaged by a particular mode''
\begin{equation}
 \vartheta_m:=\sum_{\vec x} |\psi_m({\vec x})|^2\vartheta_{\vec x}\,.
\label{eqn theta averaged}
\end{equation}
One can see in \fig\ref{sparsermtlocfigure} bottom, that the low modes are indeed localized to ``islands'' of low $\vartheta$, which are equivalent to low Polyakov loops. Hence this important effect is shared by our random matrix model, too.

\section{Conclusions} 
  \label{sect conc}

In the present paper we found a possible explanation for the emergence of
localized Poissonian modes at the low end of the high temperature QCD Dirac
spectrum. We showed that the localized modes are strongly correlated with
large fluctuations of the Polyakov loop. This lowers the effective
Matsubara frequencies for modes concentrated there. We argued that the lowest part of the Dirac spectrum
consists of this type of eigenmodes. We verified this picture for eigenmodes
of both the staggered and the overlap Dirac operator. As a side result we also
demonstrated that the spatial structure of the lowest overlap and staggered
modes is highly correlated. This shows that different discretizations of the
Dirac operator are sensitive to the same type of gauge field
fluctuations. 

We also looked at the topological charge fluctuations as
given by the index of the overlap. Assuming that at high temperature
topological objects form a dilute gas and are uncorrelated,
 we could safely rule them out in creating the localized low modes.

Finally, we proposed a dimensionally reduced random matrix
model. It is based on sparse matrices encoding the three dimensional nature of the problem through
nearest neighbor couplings from a lattice Dirac operator. To the best of our knowledge this is an example of a new kind of random matrix models for QCD, where so far only full matrices have been used. Beside chirality and the spectral gap our model reproduces the transition from
localized Poissonian to delocalized random matrix type modes observed in the
lattice Dirac spectrum as well as the correlation of the localized modes to ``islands'' of low on site ``potential''. 

It is instructive to compare the Dirac operator to the
Hamiltonian of Anderson type models, see \cite{Lee:1985zzc,Evers:2008zz} for reviews. 
In the latter case usually diagonal (on
site) disorder is responsible for creating the transition to localized
eigenmodes. In the case of the Dirac operator, the on site terms do not seem
to be relevant. In fact, in the staggered operator they are exactly
zero. However, in our dimensionally reduced three dimensional effective model
the non-zero fluctuating on site terms are dynamically generated by
fluctuations of the local Matsubara frequency resulting from fluctuations of
the Polyakov loop. In this way our dimensionally reduced effective random
matrix model is analogous to the Anderson model. It would be interesting to
study further how the presence and details of the transition depend on the
parameters and matrix size defining the random matrix ensemble.

\section{Acknowledgements}
We thank helpful discussions to Jacques Bloch, Antonio M.\ Garcia-Garcia, Ferenc Pittler, Mithat \"Unsal  and Tilo Wettig.
FB and SS are supported by DFG (BR 2872/4-2) and TGK by EU Grant
(FP7/2007-2013)/ERC n$^o$208740.


\end{document}